\documentstyle[preprint,aps,prd,eqsecnum,tighten,amsfonts,amssymb,epsfig]{revtex} 
\newcommand{\xp}{\text{$x^{\prime}$}}

\newcommand{\Mpl}{\text{$M_{\text{{\tiny P}}}$}}

\newcommand{\phip}{\text{$\phi_{+}$}}
\newcommand{\phim}{\text{$\phi_{-}$}}

\newcommand{\phipm}{\text{$\phi_{\pm}$}}

\newcommand{\Jpm}{\text{$J_{\pm}$}}
\newcommand{\phih}{\text{$\hat{\phi}$}}
\newcommand{\phihp}{\text{$\hat{\phi}_{+}$}}
\newcommand{\phihm}{\text{$\hat{\phi}_{-}$}}
\newcommand{\phihpm}{\text{$\hat{\phi}_{\pm}$}}
\newcommand{\vphi}{\text{$\varphi$}}
\newcommand{\vecphi}{\text{$\vec{\phi}$}}
\newcommand{\vphip}{\text{$\varphi_{+}$}}
\newcommand{\vphim}{\text{$\varphi_{-}$}}

\newcommand{\sqmg}{\text{$\sqrt{-g}$}}
\newcommand{\sqmgp}{\text{$\sqrt{-g'}$}}

\begin{document}
\draft
\preprint{UMDPP\#97-083}
\title{O$(N)$ Quantum fields in curved spacetime}
\author{S. A. Ramsey\thanks{Electronic address: 
{\tt sramsey@physics.umd.edu}} and 
B. L. Hu\thanks{Electronic address: {\tt hu@umdhep.umd.edu}}}
\address{Department of Physics, University of Maryland,
College Park, Maryland 20742-4111}
\date{\today}
\maketitle
\begin{abstract}
For the O$(N)$ field theory with $\lambda \Phi^4$ self-coupling,
we construct the two-particle-irreducible (2PI), closed-time-path (CTP)
effective action in a general curved spacetime.
From this we derive a set of coupled equations for the mean field and the
variance. They are  useful for studying the nonperturbative, nonequilibrium 
dynamics of a quantum field when full back reactions of the quantum field
on the curved spacetime, as well as the fluctuations on the mean field, are
required. Applications to phase transitions in the early Universe such as 
at the Planck scale or in the reheating phase of chaotic inflation are under
investigation. 
\end{abstract}
\newpage


\section{Introduction}

One major direction of research on quantum field theory in curved spacetime
\cite{dewitt:1975a,birrell:1982a,fulling:1989a} since the 1980s has been the 
application of interacting 
quantum fields to the consideration of symmetry breaking and phase transitions
in the early Universe, from the Planck to the grand unified energy scales
\cite{toms:1980a,ford:1982a,shore:1980a,vilenkin:1982a,vilenkin:1983a,allen:1983a,anderson:1985c,ratra:1985a,mazenko:1986a}.
In a series of work, Hu, O'Connor, Shen, Sinha, and Stylianopoulos
\cite{hu:1983b,oconnor:1983a,shen:1985a,hu:1986a,hu:1986c,hu:1986b,hu:1987b,sinha:1988a,stylianopoulos:1989a} 
systematically investigated the effect of spacetime curvature, dynamics, and 
finite temperature in causing a symmetry restoration of interacting quantum 
fields in curved spacetime. In general one wants to see how quantum 
fluctuations $\varphi$ around a mean field $\hat \phi$ change as a function of 
these parameters. For this purpose, the two-particle-irreducible (2PI) 
effective action was constructed for an $N$-component scalar O$(N)$ model 
with quartic interaction \cite{oconnor:1983a,anderson:1985c,hu:1987b}. Hu 
and O'Connor \cite{hu:1987b} found 
that the spectrum of the small-fluctuation operator contains interesting 
information concerning how infrared behavior of the system depends on the 
geometry and topology. The equation for $\hat \phi$ containing contributions 
from the variance of the fluctuation field $\langle\varphi^2\rangle$ depicts 
how the mean field evolves in time. This program  explored two of the three
essential elements of an investigation of 
a phase transition \cite{hu:1986a},
the geometry and topology and the field theory and infrared behavior aspects,
but not the nonequilibrium statistical-mechanical aspect.

For this and other reasons, 
Calzetta and Hu \cite{calzetta:1987a} started exploring the
closed-time-path (CTP) or Schwinger-Keldysh formalism 
\cite{schwinger:1961a,bakshi:1963a,keldysh:1964a,zhou:1985a}, 
which is formulated with an ``in-in'' 
boundary condition. Because the CTP effective action produces a real
and causal equation of motion \cite{dewitt:1986a,jordan:1986a}, 
it is well suited for 
particle production back-reaction problems 
\cite{hartle:1979a,hartle:1980a,calzetta:1989a}.  Use of the
CTP formalism in conjunction with the 2PI effective action 
\cite{cornwall:1974a} and the Wigner function \cite{wigner:1932a} enabled
Calzetta and Hu to 
construct a quantum kinetic field theory (in flat spacetime), deriving
the Boltzmann field equation from first principles \cite{calzetta:1988b}. 
The necessary ingredients
were then in place for an analysis of nonequilibrium phase transitions
\cite{calzetta:1989b}.
In recent years these tools (CTP, 2PI) have indeed been applied to the
problems of heavy-ion collisions, pair production in strong electric fields
\cite{cooper:1994a}, disoriented chiral condensates  
\cite{boyanovsky:1994b,boyanovsky:1997a}, and reheating in inflationary 
cosmology \cite{boyanovsky:1995e}.  However, none of these recent works has 
included curved spacetime effects in a self-consistent manner, where the 
spacetime governs the evolution of a quantum field and is, in turn, governed 
by the quantum field dynamics.  This is especially important for Planck scale 
processes involving quantum fluctuations with back reaction, such as
particle creation \cite{calzetta:1994a}, galaxy formation 
\cite{calzetta:1995a}, preheating, and thermalization in chaotic inflation 
\cite{boyanovsky:1996a,boyanovsky:1996c}.

With this paper we return to the problems begun by Calzetta, Hu and 
O'Connor
a decade ago. We wish to derive the coupled equations for the evolution
of the mean field and the variance for the O$(N)$ model in curved spacetime,
which should provide a solid and versatile platform for studies of phase 
transitions in the
early Universe.  The first order of business is to construct
the CTP-2PI effective action in a general curved spacetime.  The evolution
equations are derived from it. We must also deal with the divergences
arising in it.  From the vantage point of the correlation hierarchy (and the
associated master effective action) as applied to a nonequilibrium quantum
field \cite{calzetta:1995b}, there is {\em a priori\/} no reason why one 
should stop at the 2PI effective action.  Indeed, the 2PI effective action
corresponds to a further approximation from the two-loop truncation
of the master effective action constructed from the full Schwinger-Dyson
hierarchy \cite{calzetta:1993a,calzetta:1995b}. For problems where the mean 
field and the two-point function give an adequate description (which
is not the case near the critical point, where one has to be careful), 
the CTP-2PI effective action is sufficient. In particular, the 2PI 
effective action contains the commonly used large-$N$, time-dependent 
Hartree-Fock, and one-loop approximations.

The O$(N)$ model has been usefully applied to a great variety of problems in
field theory and statistical mechanics \cite{eyal:1996a}. The O$(N)$ field 
theory has the advantage that it affords use of the $1/N$ expansion
\cite{cornwall:1974a,cooper:1994a}, which yields nonperturbative evolution 
equations in the regime of strong mean field (it yields local, coupled 
dynamical equations for the mean field and the mode functions of the 
fluctuation field).  Recently it has been applied to problems of 
nonequilibrium phase transitions 
\cite{boyanovsky:1996b,boyanovsky:1995a,cooper:1997b}.
In the preheating problem 
studied in the following paper \cite{ramsey:1997b}, we shall see that this is 
particularly important for chaotic inflation scenarios \cite{linde:1985a},
in which the inflaton mean-field amplitude can be on the order of the Planck 
mass at the end of the slow roll period \cite{linde:1990a,linde:1994a}. 
The $1/N$ expansion has many attractive features, as it is known to preserve
the Ward identities of the O$(N)$ model \cite{coleman:1974a}
and to yield a covariantly conserved energy-momentum tensor 
\cite{hartle:1981a}.
Furthermore, in the limit of large $N$, the quantum effective action for the
matter fields can be interpreted as a leading-order term in the expansion of 
the full (matter plus gravity) quantum effective action \cite{hartle:1981a}.

Mazzitelli and Paz \cite{mazzitelli:1989b} have studied the $\lambda \Phi^4$ 
and O$(N)$ field theories in a general curved spacetime in the Gaussian
and large-$N$ approximations, respectively.  Their approach differs from ours
in that it is based on a Gaussian factorization which does not permit 
systematic improvement either in the loop expansion or in the $1/N$ 
approximation.  In contrast, our approach is based on a closed-time-path
formulation of the correlation dynamics, and the evolution equation we obtain
for the two-point function contains a two-loop radiative damping contribution
which is not present in the large-$N$ approximation.  At leading order
in the large-$N$ approximation, our results agree with theirs, so that their
renormalization counterterms can be directly applied to the mean field and
gap equations derived here.

This paper is organized as follows.  Secs.~\ref{sec-ctpform} and
\ref{sec-cjtform} present self-contained summaries of the two essential 
theoretical methodologies employed in this study, the closed-time-path 
formalism and the two-particle-irreducible effective action.  The adaptation 
of these tools to the quantum dynamics of a $\lambda \Phi^4$ field theory 
in curved spacetime is presented in Sec.~\ref{sec-lpfcst}.  The O$(N)$
scalar field theory is treated in Sec.~\ref{sec-ontheory}.  

Throughout this paper we use units in which 
$c = 1$. Planck's constant $\hbar$ is shown explicitly 
(i.e., not set equal to 1) except in those sections where noted. 
In these units, Newton's constant is $G = \hbar \Mpl^{-2}$, where
$\Mpl$ is the Planck mass.
We work with a four-dimensional spacetime manifold, and follow
the sign conventions\footnote{In the classification
scheme of Misner, Thorne and Wheeler \cite{misner:1973a}, the 
sign convention of Birrell and Davies \cite{birrell:1982a} is 
classified as $(+,+,+)$.}  
of Birrell and Davies \cite{birrell:1982a}
for the metric tensor $g_{\mu\nu}$, 
the Riemann curvature tensor $R_{\mu\nu\sigma\rho}$, and
the Einstein tensor $G_{\mu\nu}$.  We use greek letters to 
denote spacetime indices. The beginning latin letters 
$a,b,c,d,e,f$ are used as time branch indices (see Sec.~\ref{sec-ctpform}),
and the middle latin letters $i,j,k,l,m,n$ are used as indices in the
O$(N)$ space (see Sec.~\ref{sec-ontheory}). 
The Einstein summation convention over repeated indices is employed.
Covariant differentiation is denoted with a nabla $\nabla_{\mu}$ or
a semicolon.
 
\section{Schwinger-Keldysh formalism}
\label{sec-ctpform}
The Schwinger-Keldysh or ``closed-time-path'' (CTP)
formalism is a powerful method
for deriving real and causal evolution equations for expectation values 
of quantum operators for nonequilibrium fields, i.e., for quantum systems 
where the density matrix $\bbox{\rho}$ and the Hamiltonian $H$ do not
commute, $[H,\bbox{\rho}] \neq 0$.  This can occur, for example,  
in a field theory quantized on a dynamical background spacetime 
and also in an interacting field theory with nonequilibrium initial 
conditions.  The methods discussed here are well suited to studying the 
dynamics of an open quantum system.
Excellent reviews of the Schwinger-Keldysh method are Zhou {\em et al.\/} 
\cite{zhou:1985a} as applied to nonequilibrium quantum field theory and 
Calzetta and Hu \cite{calzetta:1987a} as applied to back reaction in 
semiclassical gravity.  
In this section we briefly review the Schwinger-Keldysh 
method in the context of
an interacting scalar field theory in Minkowski space,
with vacuum boundary conditions.


Consider a scalar field $\Phi$ in Minkowski space with a $\lambda \Phi^4$
self-interaction.  Studying the semiclassical properties of the theory consists
of taking the degrees of freedom to be the classical field $\phih$ and
fluctuations $\vphi$ about the classical field configuration.  The equation of
motion for small oscillations of $\phih$ about the stable quantum-corrected 
equilibrium configuration is obtained via a variational
principle from the effective action $\Gamma[\phih]$ \cite{jackiw:1974a}.
In the conventional Schwinger-DeWitt or ``in-out'' approach 
\cite{dewitt:1975a,schwinger:1951a}, 
one couples a $c$-number source $J$ (which is a function on ${\Bbb M}^4$)
to the field $\phi$ and computes the
vacuum persistence amplitude in the presence of the source $J$.  This
amplitude has a path integral representation
\begin{equation}
Z[J] = \exp \left( \frac{i}{\hbar} W[J] \right) =
\int D\phi \exp \left[ \frac{i}{\hbar} 
\left( S^{\text{{\tiny F}}}[\phi] + 
\int \! d^{\, 4} x J(x) \phi(x) \right) \right],
\label{eq-lpfgf}
\end{equation}
where the functional integral is a sum over classical histories of
the $\phi$ field which are pure negative frequency [i.e., all
spatial Fourier modes of $\phi$ have a time dependence like
$\exp(i \omega t)$, $\omega > 0$]
in the asymptotic past and pure positive frequency [$\sim \exp(-i \omega t)$]
in the asymptotic future.\footnote{It is noted that these
boundary conditions on the functional integral are equivalent (up to
an overall normalization) to adding a small imaginary term
$-i\epsilon \phi^2$ to the classical action, where $\epsilon > 0$.}

In a general nonequilibrium setting,  such as in a curved or 
dynamical spacetime or when $\hat{\phi}$ is time dependent,
the notion of positive frequency in the asymptotic past is in general different
from that in the asymptotic future.  
Hence, the ``in'' vacuum state
$|0,\text{in}\rangle$ defined at $x^0 = -\infty$ and
the ``out'' vacuum state $|0,\text{out}\rangle$ 
defined at $x^0 = \infty$ are not necessarily equivalent.
The generating functional $Z[J]$ defined in Eq.~(\ref{eq-lpfgf}) 
is then the vacuum persistence amplitude 
\begin{equation}
\langle0,\text{out}|0,\text{in}\rangle_{\text{{\tiny $J$}}}
= \langle0,\text{out}| T \exp \left(\frac{i}{\hbar}
\int \! d^{\, 4} x J(x) \Phi_{\text{{\tiny H}}} (x) \right) |0,\text{in}
\rangle,
\end{equation}
where $\Phi_{\text{{\tiny H}}}(x)$ is the Heisenberg field operator for the 
theory without
the source $J$.  This amplitude is in general complex. 
It follows that the classical field obtained by functional differentiation
of $-i\hbar \text{ln}Z[J]$ is the matrix element
$\langle0,\text{out}|\Phi_{\text{{\tiny H}}}|0,\text{in}\rangle$ which 
will in general be complex.
In addition, the dependence of $\phih_J \equiv \delta W / \delta J$ on $J$ 
will not, in general, be causal \cite{jordan:1986a,dewitt:1986a}.
In curved spacetime, the energy-momentum tensor $\langle T_{\mu\nu} \rangle$
is obtained by functional differentiation of $W$ with respect to
$g^{\mu\nu}$, which at one loop yields a complex matrix element of
$T_{\mu\nu}(\Phi_{\text{{\tiny H}}})$ between
the ``in'' and ``out'' vacua, where $\Phi_{\text{{\tiny H}}}$ is the 
Heisenberg field
operator and $T_{\mu\nu}(\phi)$ is the classical energy-momentum tensor
for the field \cite{dewitt:1975a}.

In the closed-time-path formalism, real and causal dynamics for $\phih$
can be obtained, as well as the expectation value of the 
energy-momentum tensor.  Let $x^0 = x^0_{\star}$ be far to the future of any 
dynamics we wish to study.  
It is not necessary to assume that $\lambda = 0$ or
that the Hamiltonian is time independent at $x^0 = x^0_{\star}$.  
As in the previous ``in-out'' approach, suppose we wish to compute
the quantum-corrected equation governing the classical field $\phih$.
Let $M = \{ (x^0,\vec{x}) | -\infty \leq x^0 \leq x^0_{\star} \}$ be
the portion of Minkowski space to the past of time $x^0_{\star}$.
We start by defining a new manifold as a quotient space
\begin{equation}
{\mathcal M} = (M \times \{+,-\} )/\sim,
\label{eq-lpfctpmn}
\end{equation}
where $\sim$ is an equivalence relation defined by the following rules
\begin{eqnarray}
&& (x,+) \sim (x',+) \qquad \text{iff $x = x'$} \nonumber \\
&& (x,-) \sim (x',-) \qquad \text{iff $x = x'$}  \label{eq-lpfctper} \\
&& (x,+) \sim (x',-) \qquad \text{iff $x = x'$ and $x^0 = x^0_{\star}$.}
\nonumber
\end{eqnarray}
The manifold ${\mathcal M}$ is orientable, provided
we reverse the sign of the volume form between the $+$ and $-$
pieces of the manifold.  
It is then straightforward to generalize the usual effective
action construction to the new manifold ${\mathcal M}$.
With the volume form on ${\mathcal M}$, we can generalize the 
classical action $S^{\text{{\tiny F}}}$ to ${\mathcal M}$,
\begin{equation}
{\mathcal S}^{\text{{\tiny F}}}[\phi_{+},\phi_{-}] =
S^{\text{{\tiny F}}}[\phi_{+}] - S^{\text{{\tiny F}}}[\phi_{-}],
\label{eq-lpfctpca}
\end{equation}
where $S^{\text{{\tiny F}}}[\phi]$ is the classical action on $M$,
and $\phi_{+}$ and $\phi_{-}$ denote the $\phi$  
field on the $+$ and $-$ branches of ${\mathcal M}$,
respectively.  The spacetime integrations in the right-hand side of
Eq.~(\ref{eq-lpfctpca}) are understood to be over $M$.
In order for $\phi_{\pm}$ to be a function on 
${\mathcal M}$, we must have 
\begin{equation}
\left.\phi_{+}(x)\right|_{x^0_{\star}} = 
\left.\phi_{-}(x)\right|_{x^0_{\star}}.
\end{equation}
The generating functional of vacuum $n$-point functions
(i.e., expectation values in the $|0,\text{in}\rangle$ vacuum)
for this theory is then defined by
\begin{equation}
Z[J_{+},J_{-}] = \int_{\text{{\tiny ctp}}} D\phi_{+} D\phi_{-}
\exp \left[ \frac{i}{\hbar} \left( 
{\mathcal S}^{\text{{\tiny F}}}[\phi_{+},\phi_{-}]
+ \int_M \! d^{\, 4} x (J_{+} \phi_{+} - J_{-} \phi_{-}) \right) \right],
\label{eq-gflpf}
\end{equation}
where $J_{+}$ and $J_{-}$ are $c$-number sources on the $+$ and $-$ 
branches of ${\mathcal M}$, respectively.
The designation ``ctp'' indicates that the 
functional integrals in Eq.~(\ref{eq-gflpf}) are over all field
configurations $(\phi_{+},\phi_{-})$ such that (i) $\phi_{+} = \phi_{-}$
at the $x^0 = x^0_{\star}$ hypersurface and (ii) $\phi_{+}$ ($\phi_{-}$)
consists of pure negative (positive) frequency modes at $x^0 = -\infty$.
It is not necessary for the normal derivatives of $\phi_{+}$ and
$\phi_{-}$ to be equal at $x^0 = x^0_{\star}$.
Because the theory is free in the asymptotic past,
a positive frequency mode\footnote{Here, the choice of
vacuum boundary conditions corresponds to adding
a small imaginary part $i\epsilon (\phip^2 - \phim^2)$ to the classical
action ${\mathcal S}^{\text{{\tiny F}}}$.  Alternatively, the
boundary conditions correspond to the usual prescription
$m^2 \rightarrow m^2 - i\epsilon$ in $S^{\text{{\tiny F}}}[\phi]$,
but with $S_{\text{{\tiny ctp}}}$ now redefined as
${\mathcal S}^{\text{{\tiny F}}}[\phip,\phim] = S^{\text{{\tiny F}}}[\phip]
- S^{\text{{\tiny F}}}[\phim]^{\star}$, where $\star$ denotes complex
conjugation \cite{calzetta:1987a}.}
is a solution to the spatial-Fourier transformed
Euler-Lagrange equation for $\phi$ whose asymptotic behavior at
$x^0 = -\infty$ is $\exp(-i\omega x^0)$, for $\omega > 0$.

\begin{figure}[htb]
\begin{center}
\epsfig{file=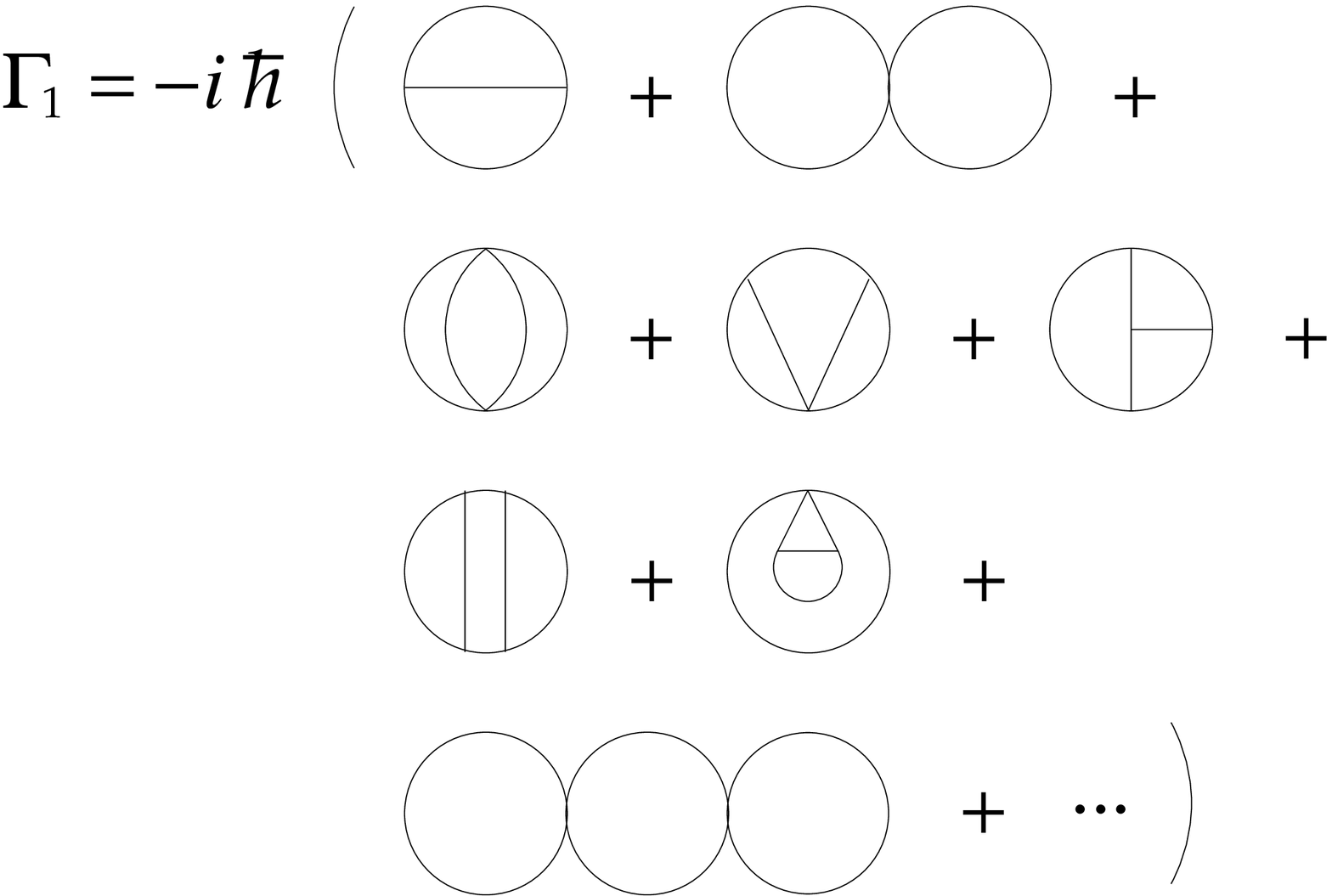,width=4.0in} 
\end{center}
\caption{Diagrammatic expansion for $\Gamma_{1}$. 
Lines represent the propagator ${\mathcal A}^{-1}_{ab}(x,\xp)$,
and vertices terminating three lines are proportional to $\phih$.
Each vertex carries spacetime $(x)$ and CTP $(+,-)$ labels.}
\label{fig-opiea}
\end{figure}

The generating functional for connected diagrams is then defined by
\begin{equation}
W[J_{+},J_{-}] = -i \hbar \text{ln} Z[J_{+},J_{-}].
\end{equation}
Classical fields on both ${+}$ and ${-}$ branches are then defined as
\begin{equation}
\phih_a(x)_{\text{{\tiny $J_{\pm}$}}} = c^{ab} 
\frac{\delta W[J_{+},J_{-}]}{\delta J_b (x)},
\label{eq-lpfph}
\end{equation}
where $a,b$ are time branch indices with index set $\{+,-\}$.
The matrix $c^{ab}$ is defined by $c^{++} = 1, \; c^{--} = -1$, and
$c^{+-} = c^{-+} = 0$.  The functional differentiation in 
Eq.~(\ref{eq-lpfph}) is carried out with variations in $\delta J_{+}$ and 
$\delta J_{-}$ which satisfy the constraint that $\delta J_{+} =
\delta J_{-}$ on the $x^0 = x^0_{\star}$ hypersurface.  The $\Jpm$ subscript
in Eq.~(\ref{eq-lpfph}) indicates the functional dependence on $\Jpm$, which 
has been shown to be causal \cite{jordan:1986a,dewitt:1986a}.
In the limit $J_{+} = J_{-} \equiv J$, the classical fields on the $+$ and
$-$ time branches become equal, 
\begin{equation}
\left.
\left(\phih_{+}(x)_{\text{{\tiny $J_{\pm}$}}}\right)
\right|_{J_{+} = J_{-} \equiv J}
 = 
\left.
\left(\phih_{-}(x)_{\text{{\tiny $J_{\pm}$}}}\right)
\right|_{J_{+} = J_{-} \equiv J} 
\equiv \phih(x)_{\text{{\tiny $J$}}} = \;
 _{\text{{\tiny $J$}} \!\!\! }\langle0,\text{in}|
\Phi_{\text{{\tiny H}}}(x)|0,\text{in}\rangle_{\text{{\tiny $J$}}},
\end{equation}
where $|0,\text{in}\rangle$ is the state which has evolved from the vacuum
at $t_0$ under the interaction $\Phi_{\text{{\tiny H}}} J$, and becomes 
the expectation
value $\langle\Phi_{\text{{\tiny H}}}\rangle$ in the limit $J = 0$.  The 
effective action is defined
via the usual Legendre transform, with $c^{ab}$ now acting as a ``metric''
on the internal $2 \times 2$ CTP field space,
\begin{equation}
\Gamma[\phihp,\phihm] = W[J_{+},J_{-}] - c^{ab} \int_M \! d^{\, 4} x J_a(x) \phih_b(x),
\label{eq-lpfead}
\end{equation}
where the $J$ subscripts on $\phih_{\pm}$ are suppressed and the functional
dependence of $J_{\pm}$ on $\phih$ via inversion of Eq.~(\ref{eq-lpfph}) is 
understood.   By direct computation, the inverse of Eq.~(\ref{eq-lpfph})
is found to be
\begin{equation}
J_a (x)_{\phihpm} = -c_{ab} \frac{\delta \Gamma[\phihp,\phihm]}{\delta 
\phih_b (x)},
\label{eq-lpfjd}
\end{equation}
where we have indicated the explicit functional dependence of $\Jpm$ 
on $\phihpm$ with a subscript, and $c_{ab}$ is the inverse of the matrix
$c^{ab}$ defined above.
In the limit $\phihp = \phihm \equiv \phih$, this yields the evolution
equation for the expectation value $_{\text{{\tiny $J$}}}\langle
\Phi_{\text{{\tiny H}}}\rangle_{\text{{\tiny $J$}}}
\equiv \phih_{\text{{\tiny $J$}}}$ in the
state which has evolved from $|0,\text{in}\rangle$ under the source interaction
$J \Phi_{\text{{\tiny H}}}$.  The evolution equation for $\phih$, the 
vacuum expectation value $\langle0,\text{in}|\Phi_{\text{{\tiny H}}}|0,
\text{in}\rangle$, is therefore
\begin{equation}
\left.
\frac{\delta \Gamma[\phihp,\phihm]}{\delta \phihp}\right|_{
\phihp = \phihm \equiv
\phih} = 
\left.-\frac{\delta \Gamma[\phihp,\phihm]}{\delta \phihm}\right|_{\phihp 
= \phihm \equiv \phih} = 0.
\label{eq-lpfctpeom}
\end{equation}
Using Eqs.~(\ref{eq-lpfjd}) and (\ref{eq-lpfead}), 
an integro-differential equation for $\Gamma$ can
be derived \cite{jordan:1986a},
in which the functional differentiations of $\Gamma$ 
with respect to $\phihpm$ are carried out with the constraint that the 
variations of $\phihpm$ satisfy $\delta \phihp = \delta \phihm$
when $x^0 = x^0_{\star}$.  The difference $\phi_a - \phih_a$ is naturally
interpreted as the fluctuations of a particular history $\phi_a$ about
the ``classical'' field configuration 
$\phih_a$.  Let us, therefore, define the 
{\em fluctuation field \/}
$\vphi_a \equiv \phi_a - \phih_a$ or, in terms of Heisenberg field operators, 
\begin{equation}
\label{eq-defff}
\vphi_{\text{{\tiny H}}} \equiv \Phi_{\text{{\tiny H}}} - 
\langle \Phi_{\text{{\tiny H}}} \rangle = \Phi_{\text{{\tiny H}}} - \phih,
\label{eq-deffluc1pi}
\end{equation}
where angular brackets around the Heisenberg field operator 
$\Phi_{\text{{\tiny H}}}$
denote an expectation value of $\Phi_{\text{{\tiny H}}}$ in the 
(time-independent)
quantum state of the system.
Performing the change of variables $\phi_a \rightarrow \vphi_a$ in the
functional integral, where
\begin{equation}
\vphi_a \equiv \phi_a - \phih_a,
\end{equation}
we obtain
\begin{equation}
\Gamma[\phihp,\phihm] = -i \hbar \text{ln} \left\{ 
\int_{\text{{\tiny ctp}}} D\vphi_{+} D\vphi_{-} \exp \left[ \frac{i}{\hbar} 
\left(
{\mathcal S}^{\text{{\tiny F}}}[\phihp + \vphip, \phihm + \vphim] -
\frac{\delta\Gamma[\phihp,\phihm]}{\delta\phih_a}\vphi_a
\right)\right]\right\}.
\label{eq-eafie}
\end{equation}
This functional integro-differential equation has a formal solution
\cite{jackiw:1974a}
\begin{equation}
\Gamma[\phihp,\phihm] = {\mathcal S}^{\text{{\tiny F}}}[\phihp,\phihm]
- \frac{i \hbar}{2} \text{ln}\,\text{det}({\mathcal A}^{-1}_{ab}) + 
\Gamma_{1}[\phihp,\phihm],
\label{eq-opieasol}
\end{equation}
where ${\mathcal A}^{ab}(x,\xp)$, the second functional derivative
of the classical action with respect to the field $\phi_{\pm}$, is
\begin{equation}
i{\mathcal A}^{ab}(x,\xp) = 
\frac{\delta^2 {\mathcal S}^{\text{{\tiny F}}}}{\delta
\phi_a(x)\phi_b(\xp)}[\phihp,\phihm].
\end{equation}
The inverse of ${\mathcal A}^{ab}$ is the one-loop propagator for the
fluctuation field $\phi$.  The functional $\Gamma_{1}$ in 
Eq.~(\ref{eq-eafie}) is defined as $-i\hbar$ times the
sum of all one-particle-irreducible vacuum-to-vacuum graphs with
propagator given by ${\mathcal A}^{-1}_{ab}(x,\xp)$ and 
vertices given by a ``shifted action''
${\mathcal S}^{\text{{\tiny F}}}_{\text{{\tiny int}}}$,
defined by
\begin{eqnarray}
{\mathcal S}^{\text{{\tiny F}}}_{\text{{\tiny int}}}[\vphip,\vphim]
= {\mathcal S}^{\text{{\tiny F}}}[\vphip + \phihp, \vphim + && \phihm] -
{\mathcal S}^{\text{{\tiny F}}}[\phihp,\phihm] - \int_M \! d^{\, 4} x \left(
\frac{\delta {\mathcal S}^{\text{{\tiny F}}}}{\delta \phi_a}[\phihpm]\right)
\vphi_a \nonumber 
\\ && - \frac{1}{2}\int_M \! d^{\, 4} x \int_M \! d^{\, 4} x' \left( 
\frac{\delta^2 
{\mathcal S}^{\text{{\tiny F}}}}{\delta \phi_a(x)\delta 
\phi_b(\xp)}[\phihpm]\right) \vphi_a(x)
\vphi_b(\xp).
\end{eqnarray}
For simplicity, we do not explicitly indicate the functional dependence of
${\mathcal S}^{\text{{\tiny F}}}_{\text{{\tiny int}}}$ on $\phihpm$.
Fig.~\ref{fig-opiea} shows the diagrammatic expansion for $\Gamma_{1}$.  
Each vertex carries a spacetime label in $M$ and a time branch label
in $\{+,-\}$.  The lowest-order contribution is order $\hbar^2$,
i.e., at two loops.  The propagator ${\mathcal A}^{-1}$ does not depend
on $\hbar$.  The $\text{ln}\det {\mathcal A}$ term in Eq.\
(\ref{eq-opieasol}) is the one-loop (order $\hbar$) term in the
CTP effective action.  The CTP
effective action, as a functional of $\phihpm$, can be computed to 
any desired order in the loop expansion using Eq.~(\ref{eq-opieasol}).
In general, this action contains divergences at each order in the
loop expansion, which need to be renormalized.

Functionally differentiating $\Gamma[\phihp,\phihm]$ with respect to
either $\phihp$ or $\phihm$ and making the identification 
$\phihp=\phihm=\phih$ [as shown in Eq.~(\ref{eq-lpfctpeom})]
yields a dynamical, real, and causal evolution equation for the
mean field $\phih$.  
Thus the 1PI effective action $\Gamma[\phihpm]$ 
yields {\em mean-field\/} dynamics for the theory, which is a lowest-order 
truncation of the correlation hierarchy 
\cite{calzetta:1993a,calzetta:1995b}.
However, for a detailed study of nonperturbative growth of quantum 
fluctuations relevant to nonequilibrium mean-field dynamics
(or a symmetry-breaking phase transition), 
it is also necessary to obtain dynamical equations for
the {\em variance\/}  of $\Phi_{\text{{\tiny H}}}$,
\begin{equation}
\langle\Phi_{\text{{\tiny H}}}^2\rangle-\langle\Phi_{\text{{\tiny H}}}
\rangle^2 = 
\langle\Phi_{\text{{\tiny H}}}^2\rangle-\phih^2 =
\langle \vphi_{\text{{\tiny H}}}^2 \rangle 
\equiv \hbar G_{++}(x,x),
\label{eq-lpf2pitpf}
\end{equation}
where $\hbar G_{++}(x,\xp)$ is the
time-ordered Green function for the fluctuation field 
$\vphi_{\text{{\tiny H}}}$, $\langle T (\vphi(x)_{\text{{\tiny H}}} 
\vphi(\xp)_{\text{{\tiny H}}}) \rangle$.
A higher-order truncation of the correlation hierarchy is 
needed in order to explicitly follow the growth of quantum fluctuations; the
2PI effective action, to which we now turn, serves this purpose.

\section{Two-Particle-Irreducible formalism}
\label{sec-cjtform}
In a nonperturbative study of nonequilibrium field dynamics in the
regime where quantum fluctuations are significant, the 1PI
effective action is inadequate because it does not permit a 
derivation of the evolution equations for the mean field 
$\langle \Phi_{\text{{\tiny H}}} \rangle$ {\em and\/} 
variance $\langle \vphi_{\text{{\tiny H}}}^2 \rangle$, at a
{\em consistent\/} order in a nonperturbative expansion scheme.
In addition, the initial data for the mean field $\phih$ do not contain
any information about the quantum state for fluctuations $\vphi$
around the mean field. 
The two-particle-irreducible (2PI) effective action method can be used 
to obtain nonperturbative dynamical equations for both 
the mean field $\phih(x)$ and two-point function $G(x,y)$,
which contains the variance, as shown in Eq.~(\ref{eq-lpf2pitpf}).
The 2PI method generalizes the 1PI effective action 
to an action $\Gamma[\phih,G]$ which 
is a functional of possible histories for both $\phih$ and $G$.  
Alternatively, the 2PI effective action can be viewed as a truncation of the
master effective action to second order in the correlation hierarchy
\cite{calzetta:1995b}. In this section we briefly review how the 
2PI method works; more thorough presentations can be
found in \cite{cornwall:1974a,calzetta:1988b}.

Unlike the 1PI method where the mean field is fixed to
be $\phih$, the 2PI method fixes the mean field to be $\phih$ and
the sum of all self-energy diagrams to be $G$. This drastically reduces
the number of
independent diagrams which must be computed in order to obtain
$\Gamma[\phih,G]$ \cite{calzetta:1993a}.  Coupled dynamical
equations for the evolution of $\phih$ and $G$ are obtained by 
separately varying $\Gamma[\phih,G]$ with respect to $G$ and $\phih$.  
Imposing
$\delta\Gamma /\delta \phih = 0$ yields an evolution equation for the mean
field $\phih$,
and setting $\delta \Gamma/\delta G = 0$ yields an evolution equation
for $G$, the ``gap'' equation.  The variance 
$\langle\vphi_{\text{{\tiny H}}}^2\rangle$
is the coincidence limit of the 
two-point function $\hbar G$, as seen from  Eq.~(\ref{eq-lpf2pitpf}).
In a nonequilibrium setting, the closed-time-path method should be
used in conjunction with the 2PI formalism in order to 
obtain real and causal dynamics for $\phih$ and $G$ 
\cite{calzetta:1988b,calzetta:1989b,calzetta:1993a}.  

Let us apply the 2PI method to a scalar $\lambda \phi^4$ theory
in Minkowski space, with vacuum initial conditions.
In a direct generalization of Sec.~\ref{sec-ctpform},
both a local source $J_a(x)$ and nonlocal source $K_{ab}(x,\xp)$ 
(which are $c$-number functions on ${\mathcal M}$) are coupled
to the field via  $\hbar c^{ab} J_a \phi_a$ and $\hbar
c^{ab} c^{cd} K_{ac}(x,\xp) \phi_b(x) \phi_d(\xp)$ interactions.  Following
Eq.~(\ref{eq-gflpf}), the CTP generating functional is defined as a 
vacuum persistence amplitude in the presence of the sources $J$ and $K$,
which has the path integral representation
\begin{eqnarray}
Z[J,K] = \int_{\text{{\tiny ctp}}} D\phi_{+} D\phi_{-}
\exp \Biggl[ \frac{i}{\hbar} \biggl( 
&& {\mathcal S}^{\text{{\tiny F}}}
[\phi_{+},\phi_{-}] + \int_M \! d^{\, 4} x c^{ab} J_a \phi_b \nonumber \\ &&
+ \frac{1}{2} \int_M \! d^{\, 4} x \int_M \! d^{\, 4} x' c^{ab} c^{cd} K_{ac}
(x,\xp)
\phi_b(x) \phi_d(\xp) \biggr) \Biggr].
\label{eq-lpf2pigf}
\end{eqnarray}
Here, ${\mathcal S}^{\text{{\tiny F}}}$ is as defined in 
Eq.~(\ref{eq-lpfctpca}), and we are using $Z[J,K]$ as a shorthand for 
$Z[J_{+},J_{-};K_{++},K_{--},K_{+-},
K_{-+}]$.
The generating functional for normalized $n$-point functions
(connected diagrams) is defined by 
\begin{equation}
W[J,K] = -i\hbar \text{ln} Z[J,K].
\label{eq-wgf2pi}
\end{equation}
The ``classical'' field $\phih_a(x)_{\text{{\tiny $JK$}}}$ and two-point 
function 
$G_{ab}(x,\xp)_{\text{{\tiny $JK$}}}$ are then given by
\begin{mathletters}
\begin{eqnarray}
\phih_a(x)_{\text{{\tiny $JK$}}} &=& c_{ab} \frac{\delta W[J,K]}{
\delta J_b(x)},
\label{eq-lpf2piphidef} 
\\
\hbar G_{ab}(x,\xp)_{\text{{\tiny $JK$}}} &=& 2 c_{ac} c_{bd} \frac{\delta 
W[J,K]}{
\delta K_{cd}(x,\xp)} - \phih_a(x)_{\text{{\tiny $JK$}}}\phih_b(\xp)_{
\text{{\tiny $JK$}}}, 
\end{eqnarray}
\label{eq-lpf2piphidefb}
\end{mathletters}
where we use the subscript $JK$ to indicate that $\phih_a$ and
$G_{ab}$ are functionals of the sources $J$ and $K$.  

In the limit $K = J = 0$, the classical field $\phih_a$ satisfies
\begin{equation}
(\phihp)_{J=K=0} = 
(\phihm)_{J=K=0} = \langle \phi | \Phi_{\text{{\tiny H}}} | \phi \rangle
\equiv \phih;
\end{equation}
i.e., it becomes the expectation value of the Heisenberg field operator 
$\Phi_{\text{{\tiny H}}}$ in the quantum state $|\phi\rangle$ (the mean field).
In the same limit, the two-point function $G_{ab}$ is the 
CTP propagator for the fluctuation field defined by
Eq.~(\ref{eq-deffluc1pi}).
The four components of the CTP propagator are, for $J = K = 0$,
\begin{mathletters}
\begin{eqnarray}
\hbar G_{++}(x,x')_{|J=K=0} &=& \langle \phi |
T\left(\vphi_{\text{{\tiny H}}}(x)
\vphi_{\text{{\tiny H}}}(x')\right) | \phi \rangle,\\
\hbar G_{--}(x,x')_{|J=K=0} &=& \langle \phi |
\tilde{T}\left(\vphi_{\text{{
\tiny H}}}(x) \vphi_{\text{{\tiny H}}}(x')\right) | \phi \rangle,\\
\hbar G_{+-}(x,x')_{|J=K=0} &=& \langle \phi | \vphi_{\text{{\tiny H}}}(x')
\vphi_{\text{{\tiny H}}}(x) | \phi \rangle,\\
\hbar G_{-+}(x,x')_{|J=K=0} &=& \langle \phi | \vphi_{\text{{\tiny H}}}(x) 
\vphi_{\text{{\tiny H}}}(x') | \phi \rangle,
\end{eqnarray}
\end{mathletters}
in the Heisenberg picture.  In the coincidence limit $x' = x$, all
four components above are equivalent to the variance
$\langle \vphi_{\text{{\tiny H}}}^2 \rangle$ defined in 
Eq.~(\ref{eq-lpf2pitpf}).
Provided we can invert Eqs.~(\ref{eq-lpf2piphidef}) and 
(\ref{eq-lpf2piphidefb})
to obtain $J$ and $K$ in terms of $\phih$ and $G$, the 2PI effective
action can be defined as the double Legendre transform (in both $J$ and
$K$) of $W[J,K]$ 
\begin{eqnarray}
\Gamma[\phih,G] = W[J,K] - && \int_M \! d^{\, 4} x c^{ab} J_a(x) \phih_b(x) 
\nonumber \\ &&
- \frac{1}{2} \int_M \! d^{\, 4} x \int_M \! d^{\, 4} x' c^{ab} c^{cd} 
K_{ac}(x,\xp) [ \hbar
G_{bd}(x,\xp) + \phih_b(x) \phih_d(\xp) ].
\label{eq-lpf2pilegtrn}
\end{eqnarray}
As with $W[J,K]$, 
we are using $\Gamma[\phih,G]$ as a shorthand for
$\Gamma[\phihp,\phihm;G_{++},G_{--},G_{+-},G_{-+}]$.  The $JK$ 
subscripting of $\phih$ and $G$ is suppressed, but the functional
dependence of $\phih$ and $G$ on $J$ and $K$ through inversion of
Eqs.~(\ref{eq-lpf2piphidef}) and (\ref{eq-lpf2piphidefb}) is understood.
By direct functional differentiation of Eq.~(\ref{eq-lpf2pilegtrn}), 
the inverses of Eqs.\~(\ref{eq-lpf2piphidef}) and 
(\ref{eq-lpf2piphidefb}) are found to be
\begin{mathletters}
\begin{eqnarray}
\frac{\delta \Gamma[\phih,G]}{\delta \phih_a(x)}
&=&  -c^{ab} J_b(x)_{\phih G} - \frac{1}{2} c^{ab} c^{cd} \int_M \! d^{\, 4} x'
( K_{bd}(x,\xp)_{\phih G} + K_{db}(\xp,x)_{\phih G})\phih_c(\xp),
\label{eq-lpf2pidgam} \\
\label{eq-lpf2pidgamb} 
\frac{\delta \Gamma[\phih,G]}{\delta G_{ab}(x,\xp)} &=&
-\frac{\hbar}{2}c^{ac}c^{bd}K_{cd}(x,\xp)_{\phih G},
\end{eqnarray}
\end{mathletters}
where the subscript ``$\phih G$'' indicates that $K$ and $J$ are functionals
of $\phih$ and $G$.  Once $\Gamma[\phih,G]$ has been calculated,
the evolution equations for $\phih$ and $G$ are given by
\begin{mathletters}
\begin{eqnarray}
\left.
\frac{\delta \Gamma[\phih,G]}{\delta \phih_a(x)}\right|_{
\phihp = \phihm \equiv 
\phih} &=& 0,
\label{eq-lpf2pieom} \\
\left.\frac{\delta \Gamma[\phih,G]}{\delta G_{ab}(x,y)}\right|_{ 
\phihp = \phihm \equiv \phih} &=& 0.
\label{eq-lpf2pieomb} 
\end{eqnarray}
\end{mathletters}
Of course, the two equations contained in 
Eq.~(\ref{eq-lpf2pieom}) are not independent,
just as in Eq.~(\ref{eq-lpfctpeom}).  In addition, only two of
equations (\ref{eq-lpf2pieomb}) are independent, one
on the diagonal and one off diagonal in the ``internal'' CTP space.
Using both Eq.~(\ref{eq-lpf2pilegtrn}) and 
Eq.~(\ref{eq-lpf2pigf}), an equation for $\Gamma[\phih,G]$ in
terms of the sources $K$ and $J$ can be derived,
\begin{eqnarray}
\Gamma[&&\phih,G] = -i \hbar \text{ln} 
\Biggl\{ \int_{\text{{\tiny ctp}}} D\phip D\phim \exp
\biggl[ \frac{i}{\hbar} \Bigl( {\mathcal S}^{\text{{\tiny F}}}[\phip,\phim]
+ c^{ab} \int_M \! d^{\, 4} x J_a(x) [\phi_b(x) - \phih_b(x)] \nonumber 
\\ && +
\frac{1}{2} c^{ac} c^{bd} \int_M \! d^{\, 4} x \int_M \! d^{\, 4} x' 
K_{ab}(x,\xp) [ \phi_c(x)\phi_d(\xp)
- \phih_c(x)\phih_d(\xp) - \hbar G_{cd}(x,\xp)]\Bigr)\biggr]\Biggr\}.
\label{eq-lpf2pigam1}
\end{eqnarray}
The sources $K$ and $J$ in the right-hand side of Eq.~(\ref{eq-lpf2pigam1})
are functionals of $\phih$, through Eqs.~(\ref{eq-lpf2pidgam}) and
(\ref{eq-lpf2pidgamb}).
Expressing this functional dependence, we 
obtain a functional integro-differential equation for $\Gamma$,
\begin{eqnarray}
\Gamma[\phih,G] = && \int_M \! d^{\, 4} x \int_M \! d^{\, 4} x' \frac{\delta 
\Gamma[\phih,G] }{
\delta G_{ba}(\xp,x)} G_{ab}(x,\xp) \nonumber \\ && 
- i \hbar \text{ln} \Biggl\{ \int_{\text{{\tiny ctp}}} D\phip D\phim \exp
\biggl[ \frac{i}{\hbar} \Bigl( {\mathcal S}^{\text{{\tiny F}}}[\phi_{+},
\phi_{-}] 
- \int_M \! d^{\, 4} x \frac{\delta \Gamma[\phih,G]}{\delta \phih_a} 
(\phi_a - \phih_a) \nonumber \\ &&
- \frac{1}{\hbar}\int_M \! d^{\, 4} x\int_M \! d^{\, 4} x' \frac{\delta 
\Gamma[\phih,G]}{\delta
G_{ba}(\xp,x)} [\phi_a(x) - \phih_a(x)][\phi_b(\xp) - \phih_b(\xp)]\Bigr)
\biggr] \Biggr\}.
\end{eqnarray}
We have dropped the $JK$ subscripting because the functional 
derivatives in the equation are only with respect to $\phih$ and $G$.
As in Sec.~\ref{sec-ctpform}, a change of variables
$D\phi_{\pm} \rightarrow D\vphi_{\pm}$ is carried out in the 
functional integral.
The resulting equation
\begin{eqnarray}
\Gamma[\phih,G] = \int_M \! && d^{\, 4} x \int_M \! d^{\, 4} x' \frac{\delta 
\Gamma[\phih,G] }{
\delta G_{ba}(\xp,x)} G_{ab}(x,\xp) \nonumber \\ &&
- i \hbar \text{ln} \Biggl\{ \int_{\text{{\tiny ctp}}} D\vphip D\vphim \exp
\biggl[ \frac{i}{\hbar} \Bigl( {\mathcal S}^{\text{{\tiny F}}}[\vphip +
\phihp, \vphim + \phihm] \nonumber \\ && - 
\int_M \! d^{\, 4} x \frac{\delta \Gamma[\phih,G]}{\delta \phih_a} \vphi_a
- \frac{1}{\hbar}\int_M \! d^{\, 4} x\int_M \! d^{\, 4} x' \frac{\delta 
\Gamma[\phih,G]}{\delta
G_{ba}(\xp,x)} \vphi_a(x) \vphi_b(\xp) \Bigr)
\biggr] \Biggr\}
\label{eq-lpf2pifideg}
\end{eqnarray}
has the formal solution \cite{cornwall:1974a}
\begin{eqnarray}
\Gamma[\phih,G] = {\mathcal S}^{\text{{\tiny F}}}[\phih] -
\frac{i \hbar}{2} \text{ln}\,\text{det}&& 
(G_{ab}) 
\nonumber \\ &&
+ \frac{i \hbar}{2} \int_M \! d^{\, 4} x \int_M \! d^{\, 4} x' 
{\mathcal A}^{ab}(\xp,x) G_{ab}(x,\xp) 
+ \Gamma_2 [\phih,G],
\end{eqnarray}
where ${\mathcal A}^{ab}$ is the second functional derivative of the
classical action ${\mathcal S}^{\text{{\tiny F}}}$,
evaluated at $\phih_a$.  The functional 
$\Gamma_2$ is $-i  \hbar$ times the sum of all two-particle-irreducible
vacuum-to-vacuum diagrams with lines given by $G_{ab}$ and vertices
given by a shifted action 
${\mathcal S}_{\text{{\tiny int}}}^{\text{{\tiny F}}}$.  We have,
for ${\mathcal A}^{ab}$,
\begin{equation}
i {\mathcal A}^{ab}(x,\xp) = \frac{\delta^2 
{\mathcal S}^{\text{{\tiny F}}}}{
\delta \phi_a(x)\phi_b(\xp)}[\phih] 
\end{equation}
and, for ${\mathcal S}^{\text{{\tiny F}}}_{\text{{\tiny int}}}$,
\begin{eqnarray}
{\mathcal S}_{\text{{\tiny int}}}^{\text{{\tiny F}}}[\vphi] =
{\mathcal S}^{\text{{\tiny F}}}[\vphi + \phih] 
- && {\mathcal S}^{\text{{\tiny F}}}[\phih] - 
\int_M \! d^{\, 4} x \left( \frac{\delta   
{\mathcal S}^{\text{{\tiny F}}}}{\delta \phi_a}[\phih] \right)
\vphi_a \nonumber \\ && 
- \frac{i}{2} \int_M \! d^{\, 4} x \int_M \! d^{\, 4} x' {\mathcal A}^{ab}
(x,\xp) \vphi_a(x) \vphi_b(\xp).
\end{eqnarray}
The shifted action for the $\lambda \phi^4$ scalar field theory is
\begin{equation}
{\mathcal S}_{\text{{\tiny int}}}^{\text{{\tiny F}}}[\vphi] =
S^{\text{{\tiny F}}}_{\text{{\tiny int}}}[\vphip] - 
S^{\text{{\tiny F}}}_{\text{{\tiny int}}}[\vphim],
\end{equation}
in terms of
\begin{equation}
S^{\text{{\tiny F}}}_{\text{{\tiny int}}}[\vphi] =
-\frac{\lambda}{6} \int_M \! d^{\, 4} x
\left( \frac{1}{4} \vphi^4  + \phih \vphi^3 \right),
\label{eq-lpf2pisa} 
\end{equation}
where the functional dependence of ${\mathcal S}^{\text{{\tiny F}}}_{\text{{
\tiny int}}}$ on $\phihpm$ is not shown explicitly.
Two types of vertices appear
in Eq.~(\ref{eq-lpf2pisa}): a vertex
which terminates four lines and a vertex terminating three lines which
is proportional to the mean field $\phih$.
The expansion for $\Gamma_2$ in terms of $G$ and $\phih$ is depicted 
graphically up to three-loop order in 
Fig.~\ref{fig-gamma2}.
Each vertex carries a spacetime label in $M$ and time branch label 
in $\{ +,- \}$.
In general, the 2PI effective action contains divergences at each order
in the loop expansion.  
It has been shown formally that if the field theory is 
renormalizable in the ``in-out'' formulation, then the ``in-in'' equations
of motion are renormalizable \cite{calzetta:1988b}.  
In the closed-time-path 
formalism it is easier to carry out explicit renormalization in the
equations of motion, i.e., the mean-field and gap equations,
which we will do in Sec.~\ref{sec-onrenorm}.

Various approximations to the full quantum dynamics can be obtained
by truncating the diagrammatic expansion for $\Gamma_2$.  Throwing
away $\Gamma_2$ in its entirety would yield the one-loop approximation.
In Fig.~\ref{fig-gamma2}, there are two two-loop diagrams, the 
``double bubble'' and the ``setting sun.''  Retaining just the double-bubble
diagram yields the time-dependent Hartree-Fock approximation 
\cite{cornwall:1974a}.  Retaining both diagrams gives a
two-loop approximation to the theory.\footnote{A different approximation,
the $1/N$ expansion, is used in Sec.~\ref{sec-ontheory} to study
the nonequilibrium dynamics of the O$(N)$ field theory.} 
  This approximation will yield a
non-time-reversal-invariant mean-field equation above threshold,
due to the setting sun diagram \cite{calzetta:1995b}. 
The time-reversal noninvariance of the mean-field equation generated by 
the 2PI effective action is a consequence of the fact that
the 2PI effective action really corresponds to a further approximation 
from the two-loop truncation (in the sense of topology of vacuum graphs) of
the {\em master\/} effective action \cite{calzetta:1995b}.  
The two-loop truncation of the master 
effective action is a functional $\Gamma_{l=2}[\phih,G,C_3]$
which depends on the mean field $\phih$, the two-point function $G$, and
the three-point function $C_3$.  The four-point function
$C_4$ also appears, but is not dynamical
due to a constraint.   The full set of equations,
\begin{mathletters}
\begin{eqnarray}
\frac{\delta\Gamma_{l=2}[\phih,G,C_3]}{\delta \phih_a} &=& 0, \\
\frac{\delta\Gamma_{l=2}[\phih,G,C_3]}{\delta G_{ab}} &=& 0, \\
\frac{\delta\Gamma_{l=2}[\phih,G,C_3]}{\delta (C_3)_{abc}} &=& 0,
\label{eq-lpf2pimtrn} 
\end{eqnarray}
\end{mathletters}
is time-reversal invariant.  However, the 2PI effective action is obtained
by solving Eq.~(\ref{eq-lpf2pimtrn}) with {\em a given choice of causal
boundary conditions\/} and substituting the resulting $C_3$ into 
$\Gamma_{l=2}$, to obtain $\Gamma_{\text{2}}[\phih,G]$.  This ``slaving''
of $C_3$ to $\phih$ and $G$ with a particular choice of boundary conditions
is what breaks the time-reversal invariance of the theory 
\cite{calzetta:1995b}.  In the paper following \cite{ramsey:1997b}
where we discuss the
preheating dynamics, we work with further
approximations which discard the setting sun diagram, and thus regain
time-reversal-invariant equations.

\section{$\lambda \Phi^4$ field theory in curved spacetime}
\label{sec-lpfcst}

In this section the quantum dynamics of a scalar $\lambda \Phi^4$ field 
theory is formulated in semiclassical gravity, where the matter fields are
quantized on a curved classical background spacetime.\footnote{
The semiclassical approximation 
is consistent with a truncation of the quantum effective action for matter 
fields and gravity perturbations at one loop 
[i.e., at order $O(\hbar)$] \cite{hartle:1979a}
or [in the case of the O$(N)$ field theory studied here] at leading order 
in the $1/N$ expansion \cite{hartle:1981a,hu:1987b}.}
The two-particle-irreducible effective action is used in
conjunction with the CTP formalism to obtain coupled evolution equations
for the mean field $\langle \Phi_{\text{{\tiny H}}} \rangle$ and
variance $\langle \Phi_{\text{{\tiny H}}}^2 \rangle - \langle
\Phi_{\text{{\tiny H}}} \rangle^2$ in the $\lambda \Phi^4$ model which are
manifestly covariant.

Let us consider a quartically self-interacting scalar
field $\phi$ in a globally hyperbolic, curved background spacetime with
metric tensor $g_{\mu\nu}$.  The diffeomorphism-invariant classical action
for this system is
\begin{equation}
S[\phi,g^{\mu\nu}] = S^{\text{{\tiny G}}}[g^{\mu\nu}] + S^{\text{{\tiny F}}}[
\phi,g^{\mu\nu}],
\label{eq-lpfcsta}
\end{equation}
where $g^{\mu\nu}$ is the contravariant metric tensor, and 
$S^{\text{{\tiny G}}}$ and $S^{\text{{\tiny F}}}$ are the classical actions
of the gravity and scalar field sectors of the theory, respectively.
For the scalar field action, we have
\begin{equation}
S^{\text{{\tiny F}}}[\phi,g^{\mu\nu}] = -\frac{1}{2} \int \! d^{\, 4} x 
\sqmg \left[
\phi (\square + m^2 + \xi R) \phi + \frac{\lambda}{12}\phi^4\right],
\label{eq-lpfcstca}
\end{equation}
where $\xi$ is the (dimensionless) coupling constant to gravity
(necessary in order for the field theory to be renormalizable
\cite{toms:1982a}), $\square$ is 
the Laplace-Beltrami operator in terms of the covariant derivative
$\nabla_{\mu}$, and $R$ is the scalar curvature.  The constant $m$ has
units of inverse length, and the $\phi$ self-coupling $\lambda$ has units
of $1/\hbar$.
Following standard procedure in semiclassical gravity \cite{birrell:1982a},
we define the semiclassical action for gravity to be
\begin{equation}
S^{\text{{\tiny G}}}[g^{\mu\nu}] = \frac{1}{16 \pi G} \int \! d^{\, 4} x 
\sqmg \left[
R - 2 \Lambda + c R^2 + b R^{\alpha\beta} R_{\alpha\beta} +
a R^{\alpha\beta\gamma\delta} R_{\alpha\beta\gamma\delta} \right],
\label{eq-lpfcstga}
\end{equation}
where $a$, $b$, and $c$ are constants with dimensions of length squared,
$R_{\alpha\beta\gamma\delta}$ is the Riemann tensor, $R_{\alpha\beta}$ is
the Ricci tensor, $\Lambda$ is the ``cosmological constant'' (with
units of inverse length-squared), $\sqmg$ is the square root of the
determinant of $g_{\mu\nu}$,
and $G$ (with units of length divided by mass) is Newton's 
constant.  As a result of to the generalized Gauss-Bonnet theorem
\cite{chern:1962a}, the constants $a$, $b$, and $c$ are not all independent
in four spacetime dimensions; let us, therefore, set $a=0$.  Classical Einstein
gravity is obtained by setting $b = 0$ and $c = 0$.
Minimal and conformal coupling (for the $\phi$ field
to gravity) correspond to setting $\xi = 0$ and $\xi = 1/6$,
respectively.  

The motivation for including the arbitrary coupling
$\xi$ and the higher-order curvature terms $R^2$ and 
$R^{\alpha\beta}R_{\alpha\beta}$ in the classical action $S$ is that
we wish to study the semiclassical dynamics of the theory.  
In the semiclassical gravity field equation and matter field equations,
divergences arise which require a renormalization of
$b$, $c$, $G$, $\Lambda$, $m$, $\xi$, and $\lambda$
\cite{birrell:1982a}.  These quantities are understood to be bare;
their observable  counterparts are renormalized.

The classical Euler-Lagrange equation
for $\phi$ is obtained by functionally differentiating
$S^{\text{{\tiny F}}}[\phi,g^{\mu\nu}]$ with
respect to $\phi$, and setting $\delta S^{\text{{\tiny F}}}/\delta \phi = 0$,
\begin{equation}
\left(\square + m^2 + \xi R + \frac{\lambda}{6} \phi^2\right) \phi = 0.
\end{equation}
The Euler-Lagrange equation for the metric $g_{\mu\nu}$
is obtained by functional differentiation  of $S$ with respect to $g^{\mu\nu}$
(it is assumed that
the variations $\delta\phi$ and $\delta g^{\mu\nu}$ are restricted 
so that no boundary terms arise),
\begin{equation}
 G_{\mu\nu} + \Lambda g_{\mu\nu} + c \, ^{(1)}H_{\mu\nu} + 
b \, ^{(2)}H_{\mu\nu} = -8 \pi G T_{\mu\nu},
\end{equation}
where the tensors $G_{\mu\nu}$, $^{(1)}H_{\mu\nu}$, and $^{(2)}H_{\mu\nu}$
are defined by \cite{fulling:1974a,fulling:1974b}
and $T_{\mu\nu}$ is the classical energy-momentum tensor,
\begin{eqnarray}
 T_{\mu\nu} = (1 - 2\xi) \phi_{;\mu} \phi_{;\nu} + &&
\left( 2 \xi - \frac{1}{2} \right) g_{\mu\nu} g^{\rho\sigma}
\phi_{;\rho}\phi_{;\sigma} - 2\xi \phi_{;\mu\nu} \phi
+ 2 \xi g_{\mu\nu} \phi \square \phi 
\nonumber \\ && - \xi G_{\mu\nu} \phi^2 +
\frac{1}{2} g_{\mu\nu} \left( m^2+ \frac{\lambda}{12}\phi^2\right)\phi^2.
\end{eqnarray}
We are interested in the dynamics of expectation values in the semiclassical
theory, which in nonequilibrium field theory does {\em not\/} follow directly
from functional differentiation of the usual Schwinger-DeWitt or ``in-out'' 
effective action.  Instead, the
Schwinger-Keldysh formalism (reviewed in Sec.~\ref{sec-ctpform}) should
be used.  Here we discuss the implementation of the Schwinger-Keldysh
method in curved spacetime.    

The first step is to generalize the closed-time-path
(CTP) manifold ${\mathcal M}$, defined in Eq.~(\ref{eq-lpfctpmn}),
to curved spacetime.  Let $\Sigma_{\star}$ be a
Cauchy hypersurface chosen so that its past domain of dependence
\cite{wald:1984a},
$D_{-}(\Sigma_{\star})$, contains all of the dynamics we wish to study.
Let us then define the manifold (with boundary)
\begin{equation}
M \equiv D_{-}(\Sigma_{\star}).
\label{eq-cstmdef}
\end{equation}
The CTP manifold ${\mathcal M}$ is defined 
following Eq.~(\ref{eq-lpfctpmn}) as a quotient space constructed by
identification on the hypersurface $\Sigma_{\star} \subset \partial M$ as in
Eq.~(\ref{eq-lpfctpmn})
where the equivalence relation is the same as Eq.~(\ref{eq-lpfctper}) 
except that
the matching of $+$ and $-$ time branches is now done on $\Sigma_{\star}$.
We construct an orientation on ${\mathcal M}$ using the canonical volume
form from $M$, $\bbox{\epsilon}_M$,
and define the volume form on ${\mathcal M}$ to be
\begin{equation}
\bbox{\epsilon}_{{\mathcal M}} = 
\left\{
\begin{array}{cc}
\bbox{\epsilon}_M & \text{on $M\times \{+\}$}, \\
-\bbox{\epsilon}_M & \text{on $M\times \{-\}$}.
\end{array}
\right.
\end{equation}
Finally, we let $\phi$ and $g^{\mu\nu}$ be independent
on the $+$ and $-$ branches of ${\mathcal M}$, provided that
$g^{\mu\nu}_{+} = g^{\mu\nu}_{-}$ and $\phi_{+} = \phi_{-}$ on
$\Sigma_{\star}$.  In other words,
$\phi$ and $g^{\mu\nu}$ must be a scalar and
a tensor, respectively, on ${\mathcal M}$.
In terms of the volume form $\bbox{\epsilon}_M$, 
we can write a scalar field action on ${\mathcal M}$, 
\begin{equation}
{\mathcal S}^{\text{{\tiny F}}}[\phipm,g^{\mu\nu}_{\pm}] = 
S^{\text{{\tiny F}}}[\phip,g^{\mu\nu}_{+}] - 
S^{\text{{\tiny F}}}[\phim,g^{\mu\nu}_{-}], 
\label{eq-sfamcm}
\end{equation}
where $S^{\text{{\tiny F}}}[\phi]$ is given by Eq.~(\ref{eq-lpfcstca}), and
$g^{\mu\nu}_{\pm}$ is the metric tensor on the $+$ and $-$ branches
of ${\mathcal M}$.
Using Eq.~(\ref{eq-lpfcstga}) we can similarly define the gravity action 
${\mathcal S}^{\text{{\tiny G}}}$ on ${\mathcal M}$,
\begin{equation}
{\mathcal S}^{\text{{\tiny G}}}[g^{\mu\nu}_{+},g^{\mu\nu}_{-}] = 
S^{\text{{\tiny G}}}[g^{\mu\nu}_{+}] - S^{\text{{\tiny G}}}[g^{\mu\nu}_{-}],
\end{equation}
where it is understood that only configurations of $g^{\mu\nu}_{\pm}$
satisfying the constraint $g^{\mu\nu}_{+} = g^{\mu\nu}_{-}$
on $\Sigma_{\star}$ are permitted.

In semiclassical gravity the scalar field theory (with action 
$S^{\text{{\tiny F}}}$)
is quantized on a classical background spacetime, with metric
$g_{\mu\nu}$, whose dynamics is determined self-consistently 
by the semiclassical geometrodynamical field equation.  
Let us denote the Heisenberg-picture field operator for the
canonically quantized $\phi$ field by $\Phi_{\text{{\tiny H}}}$.
We wish to compute the quantum effective action
$\Gamma$ for this scalar field theory, using
the two-particle-irreducible (2PI) 
method described in Sec.~\ref{sec-cjtform}.
In terms of ${\mathcal S}^{\text{{\tiny F}}}$ (now defined on the curved 
manifold ${\mathcal M}$), we define the
2PI, CTP generating functional $Z[J,K,g^{\mu\nu}]$ as follows:
\begin{eqnarray}
Z[J,K,g^{\mu\nu}] = && \int_{\text{{\tiny ctp}}} D\phip D\phim
\exp \Biggl[ \frac{i}{\hbar} \biggl( {\mathcal S}^{\text{{\tiny F}}}[\phipm,
g^{\mu\nu}_{\pm}] + \int_M \! d^{\, 4} x \sqrt{-g_c} c^{abc} J_a \phi_b 
\nonumber \\ && +
\frac{1}{2} \int_M \! d^{\, 4} x \sqrt{-g_{a'}} \int_M \! d^{\, 4} x' 
\sqrt{-g'_{c'}} c^{aba'}
c^{cdc'} K_{ac}(x,\xp) \phi_b(x) \phi_d(\xp) \biggr) \Biggr],
\label{eq-lpfcstzf}
\end{eqnarray}
where we have written $Z[J,K,g^{\mu\nu}]$ as a shorthand for
$Z[\Jpm,K_{\pm\pm},g^{\mu\nu}_{\pm}]$. The three-index symbol $c^{abc}$ is
defined by
\begin{equation}
c^{abc} = 
\left\{
\begin{array}{cc}
1 & \text{if $a = b = c = +$}, \\
-1 & \text{if $a = b = c = -$}, \\
0 & \text{otherwise}.
\end{array}
\right.
\label{eq-cabcdef}
\end{equation}
The boundary conditions on the functional integral define the initial quantum
state (assumed here to be pure).  In this and a subsequent paper 
(in which preheating dynamics of inflationary cosmology is studied
\cite{ramsey:1997b}), we are interested in the case of a quantum
state corresponding to a nonzero mean field $\phih$, with vacuum
fluctuations around the mean field.  This entails a definition of the vacuum
state for the {\em fluctuation field,\/} defined in Eq.~(\ref{eq-defff}).
In curved spacetime in general, there does not exist a unique
Poincar\'{e}-invariant vacuum state for a quantum field
\cite{dewitt:1975a,fulling:1989a}.  For an asymptotically
free field theory, a choice of ``in'' vacuum state corresponds to a choice 
of a particular orthonormal basis of solutions of the covariant Klein-Gordon
equation with which to canonically quantize the field.

From Eq.~(\ref{eq-lpfcstzf}), we can derive the two-particle-irreducible
(2PI) effective action
$\Gamma[\phih,G,g^{\mu\nu}]$ following the method of
Sec.~\ref{sec-cjtform}, with the understanding that
$\Gamma$ now depends functionally
on the metric $g^{\mu\nu}_{\pm}$ on the $+$ and $-$ time branches.
The functional differentiations should be carried out using a covariant
generalization of the Dirac $\delta$ function to the manifold $M$ 
\cite{birrell:1982a}.
The functional integro-differential equation (\ref{eq-lpf2pifideg}) for the
CTP-2PI effective action can then be generalized to the curved spacetime $M$ 
in a straightforward fashion, modulo the curved-spacetime ambiguities in the
boundary conditions of the functional integral (\ref{eq-lpfcstzf}).

The (bare) semiclassical field equations for the variance, mean field,
and metric can then be expressed in terms of variations of
${\mathcal S}^{\text{{\tiny G}}}[g^{\mu\nu}] + \Gamma[\phih,G,g^{\mu\nu}]$
with respect to $G_{\pm\pm}$, $\phi_{\pm}$, and $g^{\mu\nu}_{\pm}$,
respectively,
followed by metric and mean-field identifications between the $+$ and $-$
time branches,
\begin{mathletters}
\label{eq-lpfsee}
\begin{eqnarray}
\left.\frac{\delta ({\mathcal S}^{\text{{\tiny G}}}[g^{\mu\nu}] + 
\Gamma[\phih,G,g^{\mu\nu}])}{\delta g^{\mu\nu}_a}\right|_{ 
\phihp = \phihm = \phih ;\;\;\;
g^{\mu\nu}_{+} = g^{\mu\nu}_{-} = g^{\mu\nu}} &=& 0,\\
\left.\frac{\delta \Gamma[\phih,G,g^{\mu\nu}]}{\delta \phih_a}\right|_{
\phihp = \phihm = \phih ;\;\;\;
g^{\mu\nu}_{+} = g^{\mu\nu}_{-} = g^{\mu\nu}} &=& 0, \\
\left.\frac{\delta \Gamma[\phih,G,g^{\mu\nu}]}{\delta G_{ab}}\right|_{ 
\phihp = \phihm = \phih ;\;\;\; 
g^{\mu\nu}_{+} = g^{\mu\nu}_{-} = g^{\mu\nu}} &=& 0.
\end{eqnarray}
\end{mathletters}
As above, CTP indices are suppressed inside functional arguments.
Eqs.~(\ref{eq-lpfsee}a), (\ref{eq-lpfsee}b), and (\ref{eq-lpfsee}c) constitute
the semiclassical approximation to the full quantum dynamics for the
system described by the classical action
$S^{\text{{\tiny G}}}[g^{\mu\nu}] + S^{\text{{\tiny F}}}[\phi,g^{\mu\nu}]$.
The semiclassical field equation (with bare parameters) for $g^{\mu\nu}$ 
is obtained directly from 
Eq.~(\ref{eq-lpfsee}a),
\begin{equation}
G_{\mu\nu} + \Lambda g_{\mu\nu} + 
c\; ^{(1)\!}H_{\mu\nu} + b\; ^{(2)\!}H_{\mu\nu} = -8 \pi G
\langle T_{\mu\nu} \rangle ,
\label{eq-lpfcstsee2}
\end{equation}
where $\langle T_{\mu\nu} \rangle$ is the (unrenormalized)  
energy-momentum tensor defined by
\begin{equation}
\langle T_{\mu\nu} \rangle = \left.\frac{2}{\sqmg} \left( \frac{\delta
\Gamma[\phih,G,g^{\mu\nu}]}{\delta g^{\mu\nu}_{+}} \right)\right|_{
\phihp = \phihm = \phih; \;\;\; g^{\mu\nu}_{+} = g^{\mu\nu}_{-} =
g^{\mu\nu}}.
\label{eq-lpfcstemt}
\end{equation}
Equation (\ref{eq-lpfcstsee2}) gives the spacetime dynamics; the
dynamics of $\phih$ and $G$ are given by the mean-field and gap equations,
(\ref{eq-lpfsee}b,c).
In Eq.~(\ref{eq-lpfcstemt}), the angle brackets denote an expectation 
value of the energy-momentum tensor (with the Heisenberg field operator 
$\Phi_{\text{{\tiny H}}}$ substituted for $\phi$ in the classical theory) 
with respect to a quantum state $|\phi\rangle$ defined by the boundary
conditions on the functional integral in Eq.~(\ref{eq-lpfcstzf}).
In four spacetime dimensions (unrenormalized) $\langle T_{\mu\nu} \rangle$
has divergences which can be absorbed by the renormalization of $G$, $\Lambda$,
$b$, and $c$ \cite{birrell:1982a}.  This renormalization should be
carried out in the field equations rather than in the
CTP effective action \cite{jordan:1986a}.

The energy-momentum tensor as defined in Eq.~(\ref{eq-lpfcstemt}) 
is obtained by variation of the 2PI effective action $\Gamma$, which is a
functional of the metric $g^{\mu\nu}_{\pm}$ on both the $+$ and $-$
time branches.  From Eq.~(\ref{eq-lpfcstzf}), it is 
possible to derive $\Gamma[\phih,G,g^{\mu\nu}]$ as an arbitrary functional of 
$g^{\mu\nu}_{+}$ and $g^{\mu\nu}_{-}$.  However, in practice it is often
easier to work in the simplified case where the metric is fixed to be the
same on both the $+$ and $-$ time branches, i.e.,
\begin{equation}
g^{\mu\nu}_{+} = g^{\mu\nu}_{-} \equiv g^{\mu\nu},
\end{equation}
in the computation of $\Gamma[\phih,G,g^{\mu\nu}]$.
Once $\Gamma[\phih,G,g^{\mu\nu}]$ (or some consistent truncation of
the full quantum effective action for ${\mathcal S}^{\text{{\tiny F}}}$) 
has been computed, it is then straightforward to 
determine how $\Gamma[\phih,G,g^{\mu\nu}]$
should be generalized to the case of an arbitrary metric on ${\mathcal M}$,
for which $g^{\mu\nu}_{+}$ and $g^{\mu\nu}_{-}$ are independent.
The bare energy-momentum tensor $\langle T_{\mu\nu}\rangle$
can then be computed using Eq.~(\ref{eq-lpfcstemt}).  
Accordingly, in Sec.~\ref{sec-ontheory}, we
fix $g^{\mu\nu}_{+} = g^{\mu\nu}_{-} \equiv g^{\mu\nu}$ 
in the calculation of $\Gamma[\phih,G,g^{\mu\nu}]$.  

The semiclassical Einstein equation is a subcase of the general
geometrodynamical field equation (\ref{eq-lpfcstsee2}), obtained (after
renormalization) by setting the renormalized $b = c = \Lambda = 0$ (assuming
no cosmological constant) \cite{birrell:1982a}:
\begin{equation}
G_{\mu\nu} = -8 \pi G \langle T_{\mu\nu} \rangle.
\label{eq-seesimp}
\end{equation}
Having shown how to derive coupled evolution equations for the mean field,
variance, and metric tensor in semiclassical gravity, we now turn our 
attention to the scalar O$(N)$ model in curved spacetime. 

\section{O$(N)$ field theory in curved spacetime}
\label{sec-ontheory}
In this section we derive coupled nonperturbative dynamical equations for 
the mean field $\phih$ and variance 
$\langle \vphi_{\text{{\tiny H}}}^2 \rangle$ 
for the minimally coupled 
O$(N)$ scalar field theory with
quartic self-interaction and unbroken symmetry.
The background spacetime dynamics is given by the
semiclassical Einstein equation. These equations take into consideration
the back reaction of quantum particle production on the mean field, and quantum
fields on the dynamical spacetime, self-consistently.  
In our model the Heisenberg-picture quantum
state $|\phi\rangle$  is a coherent state for the field 
$\Phi_{\text{{\tiny H}}}$ at the initial time $\eta_0$, 
in which the expectation value $\langle \Phi_{\text{{\tiny H}}} \rangle$    
is spatially homogeneous.  The coherent state is defined with respect
to the adiabatic vacuum constructed via matching of WKB and exact mode
functions for the fluctuation field in some asymptotic region of spacetime.

The O$(N)$ field theory has the property that a systematic expansion in powers
of $1/N$ yields a nonperturbative reorganization of the diagrammatic
hierarchy which preserves
the Ward identities order by order \cite{coleman:1974a}.  
Unlike perturbation theory in the coupling
constant, which is an expansion of the theory around the vacuum configuration,
the $1/N$ expansion entails an enhancement of the mean field by $\sqrt{N}$;
this corresponds to the opposite limit of strong mean field.  
(This is precisely the situation which can arise in chaotic inflation
at the end of the slow-roll period, where the inflaton mean field amplitude
can be as large as $\Mpl/3$ \cite{ramsey:1997b}.)
As discussed in Secs.~\ref{sec-ctpform} and \ref{sec-cjtform}, the
nonequilibrium initial conditions for the mean field as well as the
nonperturbative aspect of the dynamics requires use of both closed-time-path
and two-particle-irreducible methods.  The $1/N$ expansion can be 
achieved as a further approximation from the two-loop, two-particle-irreducible
truncation of the Schwinger-Dyson equations.

Although in this study we assume a pure state, the 2PI formalism 
is also useful for an open system calculation,
in which the mean field is defined as the trace of the product of the
reduced density matrix $\bbox{\rho}$ and the Heisenberg field operator 
$\Phi_{\text{{\tiny H}}}$,
$\text{Tr}(\bbox{\rho} \Phi_{\text{{\tiny H}}})$.
When the position-basis matrix element
$\langle \phi_1 | \bbox{\rho}(\eta_0) | \phi_2 \rangle$
can be expressed as a Gaussian functional of $\phi_1$ and $\phi_2$, the
nonlocal source $K$ can encompass the initial conditions coming from
$\bbox{\rho}(t_0)$ in a natural way \cite{calzetta:1988b}.  In order to
incorporate
a density matrix whose initial condition is beyond Gaussian order in the
position basis, one can work with a higher-order truncation of the master
effective action \cite{calzetta:1995b}.  The leading-order $1/N$
approximation is equivalent to assuming a Gaussian initial density matrix;
therefore, the 2PI effective action is adequate for our purposes.

\subsection{Classical action for the O$(N)$ theory}
\label{sec-onclass}
The O$(N)$ field theory consists of $N$ spinless fields $\vec{\phi} =
\{ \phi^i \}$, $i = 1, \ldots, N$, with an 
action which is invariant under the $N$-dimensional real orthogonal group.  
The generally covariant classical action for the O$(N)$ theory (with
quartic self-interaction) plus gravity is given by
\begin{equation}
S[ \phi^i, g^{\mu\nu} ] = S^{\text{{\tiny G}}}[g^{\mu\nu}] +
S^{\text{{\tiny F}}}[ \phi^i, g^{\mu\nu} ],
\end{equation}
where $S^{\text{{\tiny G}}}[g^{\mu\nu}]$ is defined in Eq.~(\ref{eq-lpfcstga})
for the spacetime manifold $M$ with metric $g_{\mu\nu}$,
and the matter field action 
$S^{\text{{\tiny F}}}[\phi^i,g^{\mu\nu}]$ is given by
\begin{equation}
S^{\text{{\tiny F}}}[\phi^i, g_{\mu\nu}] = -\frac{1}{2} \int_M \! d^{\, 4} x \sqmg
\left[ \vecphi \cdot (\square + m^2 + \xi R ) \vecphi + \frac{\lambda}{4 N} 
( \vecphi \cdot \vecphi )^2 \right]. \label{eq-onsm} 
\end{equation}
The O$(N)$ inner product is defined by\footnote{
In our index notation, the latin letters
$i,j,k,l,m,n$  are used to designate O$(N)$ indices 
(with index set $\{1,\ldots, N\}$), while the latin letters
$a,b,c,d,e,f$ are used below to designate CTP indices 
(with index set $\{+,-\}$).}
\begin{equation}
\vecphi \cdot \vecphi = \phi^i \phi^j \delta_{ij}.
\end{equation}
In Eq.~(\ref{eq-onsm}), 
$\lambda$ is a (bare) coupling constant with dimensions 
of $1/\hbar$, and $\xi$ is the (bare) dimensionless coupling to gravity.
The classical Euler-Lagrange
equations are obtained by variation of the action $S$ 
separately with respect to the metric tensor $g_{\mu\nu}$ and the matter 
fields $\phi^i$.
In the classical action (\ref{eq-onsm}), the O$(N)$
symmetry is unbroken.  However, the O$(N)$ symmetry can be spontaneously
broken, for example, by changing $m^2$ to $-m^2$ in $S^{\text{{\tiny F}}}$.
In the symmetry-breaking case with tachyonic mass,
the stable equilibrium configuration is found to be
\begin{equation}
\vecphi \cdot \vecphi = \frac{2 N m^2}{\lambda} \equiv v^2,
\end{equation}
which is a constant.
If we wish to study the action for oscillations about the 
symmetry-broken equilibrium
configuration, the O$(N)$ invariance of Eq.~(\ref{eq-onsm}) implies that
we can choose the minimum to be in any direction; we choose it to be
in the first, i.e., $(\phi^1)^2  = v^2$.
In terms of the shifted field $\sigma = \phi^1 - v$ and the unshifted 
fields (the ``pions'')
$\pi^i = \phi^i$, $i = 1, \ldots, N-1$, the action becomes
\begin{eqnarray}
S^{\text{{\tiny F}}}[\sigma,\vec{\pi},g^{\mu\nu}] = -\frac{1}{2} \int_M \! 
 && d^{\, 4} x \sqmg
\Biggl[ \sigma (\square + m^2 + \xi R)\sigma + \vec{\pi}\cdot(\square + m^2 
+ \xi R)\vec{\pi}
+ 2 (m^2 + \xi R) \sigma^2 
\nonumber \\ && + 2 \sqrt{\frac{\lambda}{2}} M \sigma^3  + 
2 \sqrt{\frac{\lambda}{2}} M \vec{\pi}\cdot \vec{\pi} \sigma 
+ \frac{\lambda}{4}\sigma^4 - \frac{\lambda}{2}\vec{\pi}\cdot\vec{\pi}
\sigma^2 + \frac{\lambda}{4}(\vec{\pi}\cdot\vec{\pi})^2 \Biggr].
\end{eqnarray}
One can show that the effective mass of each of the ``pions'' $\vec{\pi}$
(defined as the second derivative of the potential) is zero, due to 
Goldstone's theorem.  The theorem holds for the quantum-corrected effective
potential as well \cite{peskin:1995a}.
In this paper we study the unbroken symmetry case, in order to 
focus on the parametric amplification of quantum fluctuations; this
avoids the additional complications which arise in spontaneous symmetry
breaking, e.g., infrared divergences
\cite{hu:1987b,calzetta:1997a,calzetta:1989b}.

\subsection{Quantum generating functional}
\begin{figure}[htb]
\begin{center}
\epsfig{file=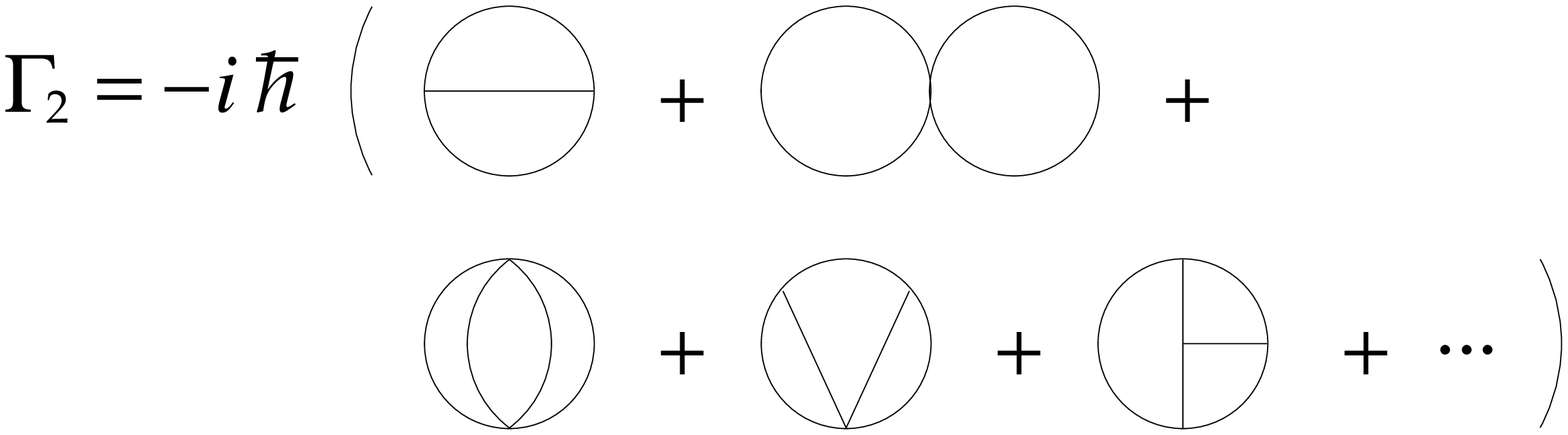,width=4.0in} 
\end{center}
\caption{Diagrammatic expansion for $\Gamma_2$. 
Lines represent the propagator $G$, and vertices are given by 
${\mathcal S}^{\text{{\tiny F}}}_{\text{{\tiny int}}}$.  The
vertices terminating three lines are proportional to $\phih$.}
\label{fig-gamma2}
\end{figure}

We aim at deriving the mean-field and gap equations at two-loop
order.  The 2PI generating functional for the O$(N)$ theory in
curved spacetime is defined using the closed-time-path method in terms of 
$c$-number sources $J^{i}_{a}$ and nonlocal $c$-number
sources $K^{ij}_{ab}$ on the CTP manifold ${\mathcal M}$,
\begin{eqnarray}
Z[J^i_a, K^{ij}_{ab}, g_{\mu\nu}] = 
&& \prod_{i,a} \int_{\text{{\tiny ctp}}} D \phi^i_a
\exp \Bigg[ \frac{i}{\hbar} \biggl( 
{\mathcal S}^{\text{{\tiny F}}}\left[ \phi, g^{\mu\nu} \right] 
+ \int_M \! d^{\, 4} x \sqmg c^{ab} \vec{J}_a \cdot \vec{\phi}_b 
\nonumber \\ &&
+ \frac{1}{2} \int_M \! d^{\, 4} x \sqmg \int_M \! d^{\, 4} x' \sqmgp c^{ab} 
c^{cd} K_{ac}^{ij}(x,\xp)
\phi^k_b(x) \phi^l_d(\xp) \delta_{ik} \delta_{jl} 
\biggr) \Biggr],
\end{eqnarray}
where the CTP classical action is defined as in Eq.~(\ref{eq-sfamcm}), with
$\phi^i_{\pm}$ replacing $\phi_{\pm}$, and the time branch indices on
$g^{\mu\nu}$ suppressed.
The sources $J^i_a$ are coupled to the field by the O$(N)$ 
vector inner product
\begin{equation}
\vec{J}_a \cdot \vec{\phi}_b = J^i_a \phi^j_b \delta_{ij}.
\end{equation}
The time branch labels on the metric tensor are suppressed for
simplicity of notation.\footnote{The suppression of CTP indices on the
metric tensor does not prevent computation of the expectation value of the
stress-energy-momentum tensor.  It will be clear how to reinstate the time
branch indices after the 2PI effective action has been explicitly
computed, e.g., in the large-$N$ approximation.}
The designation ``CTP'' on the functional integral indicates that we 
sum only over field configurations for $\phi_a^i$ on ${\mathcal M}$
which satisfy the condition $\phi^i_{+} = \phi^i_{-}$ on
$\Sigma_{\star}$, where $\Sigma_{\star}$ is defined in
Sec.~\ref{sec-lpfcst}.
In addition, the boundary conditions on the
asymptotic past field configurations for $\phi^i_{\pm}$ in the functional
integral correspond to a choice of ``in'' quantum state 
$|\phi\rangle$ for the system. 
The generating functional for normalized expectation
values is given by Eq.~(\ref{eq-wgf2pi}), with the additional functional 
dependence of both $W$ and $Z$ on $g^{\mu\nu}$ understood.
As above, we henceforth omit all indices in functional
arguments.  In terms of this functional, we can define the ``classical''
field $\phih$ and two-point function $G$ by functional differentiation,
\begin{equation}
\phih^i_a (x) = \frac{c_{ab}}{\sqmg}\frac{\delta W}{\delta J^j_b(x)}
\delta^{ij},
\label{eq-pkdef1} 
\end{equation}
\begin{equation}
\phih^i_a (x) \phih^j_b (\xp) + \hbar G^{ij}_{ab}(x,\xp) =
2\frac{c_{ac}}{\sqmg}\frac{c_{bd}}{\sqmgp} \frac{\delta W}{\delta 
K^{lm}_{cd}(x,\xp)} \delta^{ik} \delta^{jl}. 
\label{eq-pkdef2}
\end{equation}
In the zero-source limit $K^{ij}_{ab} = J^i_a = 0$,
the classical field $\phih_a^i$ satisfies
\begin{equation}
(\phih^i_{+})_{J=K=0} = (\phih^i_{-})_{J=K=0}
= \langle \phi |\Phi_{\text{{\tiny H}}}^i | \phi \rangle
\equiv \phih^i
\end{equation}
as an expectation value of the Heisenberg field operator 
$\Phi_{\text{{\tiny H}}}^i$ in the quantum state $|\phi\rangle$. 
The fluctuation field is defined [as in Eq.~(\ref{eq-deffluc1pi})]
in terms of the Heisenberg field operator $\Phi_{\text{{\tiny H}}}$
and the mean field $\phih$ (times the identity operator),
\begin{equation}
\vphi_{\text{{\tiny H}}}^i = \Phi_{\text{{\tiny H}}}^i - \phih^i.
\end{equation}
In the same limit $J = K = 0$, 
the two-point function $G_{ab}^{ij}$ becomes the CTP propagator for 
the fluctuation field.
The four components of the CTP propagator are (for $J^i_a = K^{ij}_{ab} = 0$)
\begin{mathletters}
\begin{eqnarray}
\hbar G^{ij}_{++}(x,x')_{|J=K=0} &=&
\langle \phi | T(\vphi_{\text{{\tiny H}}}^i(x)
\vphi_{\text{{\tiny H}}}^j(x')) | \phi \rangle,  \\
\hbar G^{ij}_{--}(x,x')_{|J=K=0} &=&
\langle \phi | \tilde{T}(\vphi_{\text{{
\tiny H}}}^i(x) \vphi_{\text{{\tiny H}}}^j(x')) | \phi \rangle,  \\
\hbar G^{ij}_{+-}(x,x')_{J=K=0} &=&
\langle \phi | \vphi_{\text{{\tiny H}}}^j(x')
\vphi_{\text{{\tiny H}}}^i(x) | \phi \rangle, \\
\hbar G^{ij}_{-+}(x,x')_{J=K=0} &=&
\langle \phi | \vphi_{\text{{\tiny H}}}^i(x)
\vphi_{\text{{\tiny H}}}^j(x') | \phi \rangle.
\end{eqnarray}
\end{mathletters}
In the coincidence limit $x' = x$, all
four components above are equivalent to the mean-squared fluctuations
(variance) about the mean field $\phih^i$,
\begin{equation}
\hbar G^{ii}_{++}(x,x)_{|J=K=0} =
\langle \phi | (\vphi_{\text{{\tiny H}}}^i)^2 | \phi \rangle 
= \langle (\vphi_{\text{{\tiny H}}}^i)^2 \rangle.
\label{eq-onfldef}
\end{equation}
Provided that the above equations can be inverted to give 
$J^i_a$ and $K^{ij}_{ab}$ in terms of $\phih^i_a$ and $G^{ij}_{ab}$, 
we can define the 2PI effective action as a double
Legendre transform of $W$,
\begin{eqnarray}
\Gamma[&&\phih, G, g^{\mu\nu}] = W[J, K, g^{\mu\nu}] -
\int_M \! d^{\, 4} x \sqmg c^{ab} J^i_a \phih^j_b \delta_{ij} 
\nonumber \\ && -
\frac{1}{2} \int_M \! d^{\, 4} x \sqmg \int_M \! d^{\, 4} x' \sqmgp c^{ab} 
c^{cd} K^{ij}_{ac}(x,\xp) 
\Bigl[
\hbar G^{kl}_{bd}(x,\xp) + \phih^k_b(x) \phih^l_d (\xp) \Bigr] 
\delta_{ik} \delta_{jl}, \label{eq-ea2pidef}
\end{eqnarray}
where $J^i_a$ and $K^{ij}_{ab}$ above denote the inverses of
Eqs.~(\ref{eq-pkdef1}) and (\ref{eq-pkdef2}).
From this equation, it is clear that the inverses of Eqs.~
(\ref{eq-pkdef1}) and (\ref{eq-pkdef2})
can be obtained by straightforward functional differentiation of 
$\Gamma$,
\begin{equation}
\frac{1}{\sqmg}\frac{\delta \Gamma}{
\delta \phih^i_a (x)} = c^{ab} \delta_{ij} \Bigl( -J^j_b (x) 
- \frac{1}{2} c^{cd} \int_M \! d^{\, 4} x' \sqmgp \bigl[ K^{jk}_{bd}(x,\xp) +
K^{jk}_{db} (\xp,x) \bigr] \phih^l_d \delta_{kl}\Bigr)
\label{eq-onkdef1} 
\end{equation}
\begin{equation}
\frac{1}{\sqmg} \frac{\delta \Gamma}{\delta
G^{ij}_{ab}(x,\xp)}\frac{1}{\sqmgp} =
-\frac{\hbar}{2}c^{ac}c^{bd} K^{jk}_{bd}(\xp,x).
\label{eq-onkdef2}
\end{equation}
Performing the usual field shifting involved in the background field 
approach \cite{jackiw:1974a}, it can be shown that the 2PI effective action
which satisfies Eqs.~(\ref{eq-ea2pidef}), 
(\ref{eq-onkdef1}), and (\ref{eq-onkdef2}) can be written
\begin{eqnarray}
\Gamma[\phih,G,g^{\mu\nu}] = 
&& {\mathcal S}^{\text{{\tiny F}}}[\phih,g^{\mu\nu}] 
- \frac{i \hbar}{2} \text{ln}\, 
\text{det} \left[ G^{ij}_{ab} \right] 
\nonumber \\ &&
+ \frac{i \hbar}{2} \int_M \! d^{\, 4} x \sqmg \int_M \! d^{\, 4} x' 
\sqmgp {\mathcal A}_{ij}^{ab}(\xp,x)
G^{ij}_{ab}(x,\xp) + \Gamma_2[\phih,G,g^{\mu\nu}],\label{eq-ea2piai}
\end{eqnarray}
where the kernel ${\mathcal A}$ is the second functional derivative
of the classical action with respect to the field $\phi$, 
\begin{equation}
i {\mathcal A}_{ij}^{ab}(x,\xp) = \frac{1}{\sqmg} 
\Biggl(\frac{\delta^2 {\mathcal S}^{\text{{\tiny F}}}}{
\delta \phi^i_a (x) \phi^j_b (\xp)} [\phih,g^{\mu\nu}]\Biggr) \frac{1}{\sqmgp},
\end{equation}
and $\Gamma_2$ is a functional to be defined below.
Evaluating ${\mathcal A}^{ab}_{ij}$
by differentiation of Eq.~(\ref{eq-onsm}), we find
\begin{eqnarray}
i {\mathcal A}_{ij}^{ab}(x,x') &&
= -\biggl\{\delta_{ij}c^{ab}[\square_x + m^2 + 
\xi R(x)] 
\nonumber \\ &&
+ \frac{\lambda}{2N} c^{abcd} \left[ [\phih_c^k(x) \phih_d^l(x)]\delta_{ij}
\delta_{kl} + 2 \phih^k_c(x)\phih^l_d(x)\delta_{ik}\delta_{jl}\right]\biggr\}
\delta^4(x-x')\frac{1}{\sqmgp},
\end{eqnarray}
where the four-index symbol $c^{abcd}$ is defined in exact analogy with
Eq.~(\ref{eq-cabcdef}).
In Eq.~(\ref{eq-ea2piai}), $\Gamma_2$ is $-i\hbar$ times 
the sum of all two-particle-irreducible
vacuum-to-vacuum graphs with propagator $G$ and vertices given by the 
shifted action ${\mathcal S}^{\text{{\tiny F}}}_{\text{{\tiny int}}}$, 
defined by
\begin{eqnarray}
{\mathcal S}^{\text{{\tiny F}}}_{\text{{\tiny int}}}[\vphi,g^{\mu\nu}&&] = 
{\mathcal S}^{\text{{\tiny F}}}[\vphi + \phih, g^{\mu\nu}] - 
{\mathcal S}^{\text{{\tiny F}}}[\phih, g^{\mu\nu}]
-\int_M \! d^{\, 4} x \biggl( \frac{\delta {\mathcal S}^{\text{{\tiny F}}}}{
\delta \phi^i_a} 
[\phih,g^{\mu\nu}]\biggr) \vphi^i_a \nonumber \\ &&
- \frac{1}{2} \int_M \! d^{\, 4} x \int_M \! d^{\, 4} x' \biggl( 
\frac{\delta^2 {\mathcal S}^{
\text{{\tiny F}}}
}{\delta \phi^i_a(x) \phi^j_b (\xp)}[\phih, g^{\mu\nu}]\biggr)
\vphi^i_a(x)\vphi^j_b(\xp). \label{eq-sintctp}
\end{eqnarray}
The expansion of $\Gamma_2$ in terms of $G$ and $\phih$ is depicted
graphically in Fig.~\ref{fig-gamma2}.  From
Eqs.~(\ref{eq-sintctp}) and (\ref{eq-onsm}),
${\mathcal S}^{\text{{\tiny F}}}_{\text{{\tiny int}}}$
is easily evaluated, and we find
\begin{equation}
{\mathcal S}^{\text{{\tiny F}}}_{\text{{\tiny int}}}
[\vphi,g^{\mu\nu}] =
S^{\text{{\tiny F}}}_{\text{{\tiny int}}}[\vphip,g^{\mu\nu}] - 
S^{\text{{\tiny F}}}_{\text{{\tiny int}}}[\vphim,g^{\mu\nu}],
\label{eq-sinteval}
\end{equation}
in terms of an action $S^{\text{{\tiny F}}}_{\text{{\tiny int}}}$
on $M$ defined by
\begin{equation}
S^{\text{{\tiny F}}}_{\text{{\tiny int}}}[\vphi,g^{\mu\nu}] =
-\frac{\lambda}{2 N} \int_M \! d^{\, 4} x \sqmg \left[
\frac{1}{4} (\vec{\vphi}\cdot\vec{\vphi})^2 +
(\vec{\phih}\cdot\vec{\vphi})(\vec{\vphi}\cdot\vec{\vphi})\right].
\label{eq-sintevals}
\end{equation}
The two types of vertices in Fig.~\ref{fig-gamma2} are readily
apparent in Eq.~(\ref{eq-sintevals}).  The first term corresponds to the
vertex which terminates four lines; the second term corresponds
to the vertex which terminates three lines and is proportional to
$\phih$.

The action $\Gamma$ including the full diagrammatic series for $\Gamma_2$
gives the full dynamics for $\phih$ and $G$ in the O$(N)$ theory.
It is of course not feasible to obtain an exact, closed-form expression
for $\Gamma_2$ in this model.
Various approximations to the full 2PI effective action can be obtained by
truncating the diagrammatic expansion for $\Gamma_2$.  Which approximation
is most appropriate depends on the physical problem under consideration.
\begin{enumerate}
\item
Retaining both the ``setting-sun'' and the ``double-bubble'' diagrams of
Fig.~\ref{fig-gamma2} corresponds to the two-loop, two-particle-irreducible
approximation \cite{calzetta:1995b}.  This approximation contains
two-particle scattering through the setting-sun diagram.
\item
A truncation of $\Gamma_2$ retaining only the ``double-bubble'' diagram
of Fig.~\ref{fig-gamma2} yields equations for $\phih$ and $G$ which
correspond to the time-dependent Hartree-Fock approximation to the full
quantum dynamics \cite{cornwall:1974a,cooper:1994a}.  This approximation
does not preserve Goldstone's theorem, but is
energy conserving (in Minkowski space) \cite{cooper:1994a}.  
\item
Retaining only the $(\text{tr}G^{ij}_{ab})^2$ piece of the double-bubble
diagram corresponds to taking the leading order $1/N$ approximation,
shown below in Sec.~\ref{sec-onlna}.
\item
A much simpler approximation consists of 
discarding $\Gamma_2$ altogether.  This yields 
the one-loop approximation, whose limitations
have been extensively documented in the literature
\cite{calzetta:1988b,calzetta:1993a,calzetta:1995b,calzetta:1989b}. 
\end{enumerate}

Let us first evaluate the 2PI effective action at two loops
\cite{hu:1987b,calzetta:1995a}.  This is the most general
of the various approximations described above.
Here both two-loop diagrams in
Fig.~\ref{fig-gamma2} are retained.  The 2PI effective action is given
by Eq.~(\ref{eq-ea2piai}), and in this approximation, $\Gamma_2$ is given by
%
%
\begin{eqnarray}
\Gamma_2 && [\phih,G,g^{\mu\nu}] = 
\frac{\lambda \hbar^2}{4 N} \Biggl[ -\frac{1}{2}c^{abcd} \int_M \! d^{\, 4} 
x \sqmg
[ G^{ij}_{ab}(x,x)G^{kl}_{cd}(x,x) + 2 G^{ik}_{ab}(x,x)G^{jl}_{cd}(x,x)
] \delta_{ij} \delta_{kl} \nonumber \\ &&
+ \frac{i \lambda}{N} c^{abcd} c^{a'b'c'd'} \int_M \! d^{\, 4} x \sqmg 
\int_M \! d^{\, 4} x' \sqmgp 
\phih^i_a(x) \phih^{i'}_{a'}(x') [ G^{i i'}_{bb'}(x,x') 
G^{jj'}_{cc'}(x,x')G^{kk'}_{dd'}(x,x') \nonumber \\  &&
+ 2 G^{ij'}_{bd'}(x,x') 
G^{jk'}_{cc'}(x,x')G^{ki'}_{db'}(x,x')]\delta_{jk}\delta_{j'k'} \Biggr].
\end{eqnarray}
Functional differentiation of $\Gamma$ with respect to $\phih$ and $G$ leads
to the mean-field and gap equations, respectively.  The gap equation obtained
at two loops is given by
\[
(G^{-1})^{ab}_{ij}(x,x') = {\mathcal A}^{ab}_{ij}(x,x') + 
\frac{i\lambda\hbar}{2N}c^{abcd} \delta^4(x-x') 
\left[ \delta^{ij}\delta_{kl} G^{kl}_{cd}(x,x)
 + 2 G^{ij}_{cd}(x,x) \right] 
\]
\begin{eqnarray}
+ \frac{\hbar\lambda^2}{2 N^2}  
c^{acde}c^{bc'd'e'} \delta_{kk'} \delta_{ll'} \Bigl[&&
\phih^i_c(x)\phih^j_{c'}(x') G^{kl}_{dd'}(x,x')G^{k'l'}_{ee'}(x,x') 
\nonumber \\ && +
2\phih^k_c(x)\phih^l_{c'}(x') G^{k'l'}_{dd'}(x,x')G^{ij}_{ee'}(x,x') 
\nonumber \\ && + 
2\phih^i_c(x)\phih^k_{c'}(x') G^{lj}_{dd'}(x,x')G^{l'k'}_{ee'}(x,x') 
\nonumber \\ && + 
2\phih^k_c(x)\phih^l_{c'}(x') G^{k'j}_{dd'}(x,x')G^{il'}_{ee'}(x,x') 
\nonumber \\ && + 
2\phih^k_c(x)\phih^j_{c'}(x') G^{k'l}_{dd'}(x,x')G^{jl'}_{ee'}(x,x') 
\Bigr]. \label{eq-on2lge}
\end{eqnarray}
The mean-field equation is found to be
\begin{eqnarray}
\Bigl( c^{cb} (\square + m^2 + \xi R) + &&
 c^{abcd} \frac{\lambda}{2N} \phih^i_a \phih^j_d \delta_{ij} \Bigr) \phih^m_b 
- \frac{i \hbar^2 \lambda^2}{4 N^2} \int_M d^{\,4}x' \sqmgp \Sigma^{cm}(x,x')
\nonumber \\ &&
 + \frac{\hbar \lambda c^{abcd}}{2 N} \biggl\{
\delta_{ij} \phih^m_d G^{ij}_{ab}(x,x) 
+ \delta_{jl} \delta^m_i \phih^l_d [ G^{ij}_{ab}(x,x) + G^{ji}_{ab}(x,x)]
\biggr\},
\label{eq-on2lmfe}
\end{eqnarray}
in terms of a nonlocal function $\Sigma^{cm}(x,x')$ defined by
\begin{eqnarray}
\Sigma^{em}(x,x') = c^{ebcd} c^{a'b'c'd'} \phih^i_{a'}(x') \Bigl[ && 
  G^{mi'}_{bb'}(x,x') G^{jj'}_{cc'}(x,x') G^{kk'}_{dd'}(x,x') 
\nonumber \\ && +
2 G^{mj'}_{bd'}(x,x') G^{jk'}_{cc'}(x,x') G^{ki'}_{b'd}(x,x') 
\nonumber \\ && +
G^{i'm}_{b'b}(x',x) G^{jj'}_{c'c}(x',x) G^{kk'}_{d'd}(x',x) 
\nonumber \\ && +
2 G^{i'j'}_{b'd}(x',x) G^{kk'}_{c'c}(x',x) G^{jm}_{d'b}(x',x) \Bigr] 
\delta_{jk} \delta_{j'k'} \delta_{ii'}.
\end{eqnarray}
Taking the limit $\phih^i_{+} = \phih^i_{-} = \phih^i$ in 
Eqs.~(\ref{eq-on2lge}) and (\ref{eq-on2lmfe}) 
yields coupled equations for the mean field $\phih^i$
and the CTP propagators $G^{ij}_{ab}$, on the fixed background spacetime 
$g^{\mu\nu}$.  The equations, as well as the semiclassical Einstein  
equation obtained from Eq.~(\ref{eq-lpfsee}a), 
are real and causal, and correspond
to expectation values in the $\phihp = \phihm = \phih$ limit.  
The $O(\lambda^2)$ parts of the above equations 
are nonlocal and dissipative.  The nonlocal aspect
makes numerical solution difficult;
the dissipative aspect will be addressed in a future publication
\cite{ramsey:1997e}.
One can regain the perturbative (amplitude) expansion
for the CTP effective action at two loops 
by expanding the one-loop CTP propagators in Eq.~(\ref{eq-on2lmfe})
in a functional power series in $\phih$.

\subsection{Large-$N$ approximation}
\label{sec-onlna}
We now carry out the $1/N$ expansion to obtain local, covariant,
nonperturbative  mean-field and gap equations for the O$(N)$ field theory
in a general curved spacetime.  The $1/N$ expansion is a controlled
nonperturbative approximation scheme which can be used to study 
nonequilibrium quantum 
field dynamics in the regime of strong quasiclassical field amplitude 
\cite{boyanovsky:1996b,boyanovsky:1995a,cooper:1994a,cooper:1997b}.   
In the large-$N$ approach, the 
large-amplitude quasiclassical field is modeled by $N$ fields, and the
quantum-field-theoretic generating functional is expanded in powers of
$1/N$.  In this sense the method is a controlled expansion in a small
parameter.  Unlike perturbation theory in the coupling constant, which
corresponds to an expansion of the theory around the vacuum,
the large-$N$ approximation corresponds to an expansion of the field theory
about a strong quasiclassical field configuration \cite{cooper:1994a}.
At a particular order in the $1/N$ expansion, the approximation yields
truncated Schwinger-Dyson equations which are gauge and renormalization-group 
invariant, unitary, and (in Minkowski space) energy conserving 
\cite{cooper:1994a}.  In contrast, the Hartree-Fock approximation cannot
be systematically improved beyond leading order and (in the case of
spontaneous symmetry breaking) violates Goldstone's theorem 
\cite{cooper:1997b}.

Let us implement the leading order large-$N$ approximation in
the two-loop, 2PI mean-field and gap equations (\ref{eq-on2lmfe}) 
and (\ref{eq-on2lge}) derived above.  This amounts to computing the
leading-order part of $\Gamma$ in the limit of large $N$, which is 
$O(N)$.  In the unbroken symmetry case, this is easily carried out
by scaling $\phih$ by $\sqrt{N}$ and leaving $G$ unscaled 
\cite{cornwall:1974a},
\begin{mathletters}
\begin{eqnarray}
\phih^i_a(x) & \rightarrow & \sqrt{N} \phih_a(x), \\
G^{ij}_{ab}(x,x') & \rightarrow & G_{ab}(x,x')\delta^{ij},  \\
{\mathcal A}_{ij}^{ab}(x,x') & \rightarrow & {\mathcal A}^{ab}(x,x')
\delta_{ij}, \\
\vphi^i_a(x) & \rightarrow & \vphi_a(x), 
\label{eq-onscl} 
\end{eqnarray}
\end{mathletters}
for all $i,j$.
The Heisenberg field operator $\vphi_{\text{{\tiny H}}}^i$ 
scales like $\vphi^i_a$ in Eq.~(\ref{eq-onscl}).
In the above equations, the connection between the large-$N$ limit and
the strong mean-field limit is clear.

The truncation of the $1/N$ expansion should be carried out in the 2PI
effective action, where it can be shown that the three-loop and higher-order
diagrams do not contribute (at leading order in the $1/N$ expansion).  
Let us now also allow the metric $g_{\mu\nu}$ to be specified 
independently on the $+$ and $-$ 
time branches.  We find, for the classical action,
\begin{equation}
{\mathcal S}^{\text{{\tiny F}}}[\phi,g^{\mu\nu}] =
S^{\text{{\tiny F}}}[\phi_{+},g^{\mu\nu}_{+}] - 
S^{\text{{\tiny F}}}[\phi_{-},g^{\mu\nu}_{-}], 
\end{equation}
where
\begin{equation}
S^{\text{{\tiny F}}}[\phi,g^{\mu\nu}] =
-\frac{N}{2} \int_M \! d^{\, 4} x \sqmg \left[ \phi ( \square + m^2 + 
\xi R ) \phi +
\frac{\lambda}{2} \phi^4 \right].
\end{equation} 
The inverse of the one-loop propagator is\footnote{
Note that the index $b$ is not to be summed in the right-hand side of
Eq.~(\ref{eq-onloolp}), and the $c$ subscript on $\square$ and $R$ is a CTP
index.}
\begin{equation}
i{\mathcal A}^{ab}(x,x') = -\left[ c^{abc}[\square^x_c + m^2 + \xi R_c(x)]
+ \frac{\lambda}{2} c^{abcd} \phih_c(x)\phih_d(x) \right] \delta^4(x-x')
\frac{1}{\sqmgp_b}. \label{eq-onloolp}
\end{equation}
Finally, for the CTP-2PI effective action
at leading order in the $1/N$ expansion, we obtain
\begin{eqnarray}
\Gamma[\phih,G,g^{\mu\nu}] = && {\mathcal S}^{\text{{\tiny F}}}
[\phih,g^{\mu\nu}] - \frac{i \hbar N}{2} \text{ln}\,\text{det}
\left[ G_{ab} \right] 
\nonumber \\ && + \frac{i \hbar N}{2}
\int_M \! d^{\, 4} x \sqmg_a \int_M \! d^{\, 4} x' \sqmgp_b 
{\mathcal A}^{ab}(x',x) G_{ab}(x,x')  \nonumber \\ && -
\frac{\lambda \hbar^2 N}{8} c^{abcde} \int_M \! d^{\, 4} x 
\sqmg_e G_{ab}(x,x) G_{cd}(x,x)
+ O(1).
\end{eqnarray}
Applying Eq.~(\ref{eq-lpfsee}b) and taking the limits $\phihp = \phihm =\phih$
and $g^{\mu\nu}_{+} = g^{\mu\nu}_{-} = g^{\mu\nu}$, we obtain the
gap equation for $G_{ab}$ at leading order in the $1/N$ expansion,
\begin{equation}
(G^{-1})^{ab}(x,x') = \hat{{\mathcal A}}^{ab}(x,x') 
+ \frac{i \hbar \lambda}{2}
c^{abcd} G_{cd}(x,x) \delta^4(x-x')\frac{1}{\sqmgp} + O\left(
\frac{1}{N}\right),
\label{eq-ongelm} 
\end{equation}
where
\begin{equation}
i\hat{{\mathcal A}}^{ab}(x,x') \equiv -\left[ c^{ab} [ \square + m^2 + 
\xi R(x) ]
+ \frac{\lambda}{2}c^{abcd} \phih_c(x) \phih_d(x) \right] \delta^4(x-x')
\frac{1}{\sqmgp}.
\end{equation}
Similarly, we obtain the mean-field equation for $\phih$ at leading order
in the $1/N$ expansion,
\begin{equation}
\left(\square + m^2 + \xi R + \frac{\lambda}{2} \phih^2 + \frac{\hbar 
\lambda}{2}
G_{++}(x,x) \right) \phih(x) + O\left(\frac{1}{N}\right) = 0,
\label{eq-onmflm}
\end{equation}
where we note that $G_{++}(x,x) = G_{ab}(x,x)$ for all $a,b$, 
which can be seen from Eq.~(\ref{eq-ongelm}); therefore, to get a consistent  
set of dynamical equations, we need only consider the $++$ component
of Eq.~(\ref{eq-ongelm}).  It should also be noted that $G_{ab}(x,x)$ is 
formally divergent.  Regularization of the coincidence limit of the
two-point function and the energy-momentum tensor is necessary. 
Multiplying Eq.~(\ref{eq-ongelm}) by $G$ and
integrating over spacetime, we obtain a differential equation for
the $++$ CTP Green function,
\begin{equation}
\left( \square_x + m^2 + \xi R(x) + \frac{\lambda}{2}\phih^2(x) + 
\frac{\hbar \lambda}{2} G_{++}(x,x) 
\right) G_{++}(x,x') + O\left(\frac{1}{N}\right) = 
\delta^4(x-x') \frac{-i}{\sqmgp},
\label{eq-onige}
\end{equation}
where boundary conditions must be specified on $G_{++}$.  

Equations (\ref{eq-onmflm}) and (\ref{eq-onige}) are the covariant
evolution equations for the mean field $\phih$ and the two-point function
$G_{++}$ at leading order in the $1/N$ expansion.  
Following Eq.~(\ref{eq-onfldef}), we denote the coincidence limit 
of $\hbar G_{++}(x,x)$ by
$\langle \vphi_{\text{{\tiny H}}}^2 \rangle$.
With the inclusion of the semiclassical gravity field equation
(\ref{eq-lpfcstsee2}),
these equations form a consistent, closed set of dynamical equations
for the mean field $\phih$, the time-ordered fluctuation-field
Green function $G_{++}$, and the metric $g_{\mu\nu}$.

The one-loop equations for $\phih$ and $G$ can be obtained from 
the leading-order equations by setting $\hbar=0$ in Eq.~(\ref{eq-onige}),
while leaving the mean-field equation (\ref{eq-onmflm}) unchanged.  In
the Hartree approximation, the gap equation is 
unchanged from Eq.~(\ref{eq-ongelm}), and the mean-field equation is obtained
from Eq.~(\ref{eq-onmflm}) by changing $\hbar \rightarrow 3\hbar$
\cite{boyanovsky:1995a}.  The principal limitation of the leading-order
large-$N$ approximation is that it neglects the setting-sun diagram which
is the lowest-order contribution to collisional thermalization of the
system \cite{calzetta:1988b}.  The system, therefore, does not thermalize
at leading order in $1/N$, and the approximation breaks down on a
time scale $\tau_2$ which is on the order of the mean free time for binary
scattering \cite{cooper:1997b} (collisional thermalization processes in
reheating the post-inflationary Universe is discussed in a 
subsequent paper \cite{ramsey:1997b}).

Let us now use Eq.~(\ref{eq-lpfsee}a) 
to derive the bare semiclassical Einstein
equation for the O$(N)$ theory at leading order in $1/N$.  This equation
contains two parts $\delta {\mathcal S}^{\text{{\tiny G}}}/\delta 
g^{\mu\nu}_{+}$ and $\delta \Gamma /\delta g^{\mu\nu}_{+}$.  The latter
part is related to the bare energy-momentum tensor $ \langle T_{\mu\nu}
\rangle$ by Eq.~(\ref{eq-lpfcstemt}).  At leading order in $1/N$, 
$\langle T_{\mu\nu}\rangle$ is given by a sum of classical and quantum parts,
\begin{equation}
\langle T_{\mu\nu} \rangle = T^{\text{{\tiny C}}}_{\mu\nu} +
T^{\text{{\tiny Q}}}_{\mu\nu} - \frac{\lambda N}{8} \langle \vphi_{\text{{\tiny
H}}}^2 \rangle^2 g_{\mu\nu},
\end{equation}
where we define the classical part of $\langle T_{\mu\nu}
\rangle$ by
\begin{eqnarray}
T^{\text{{\tiny C}}}_{\mu\nu} = N \Biggl[&& (1-2\xi) \phih_{;\mu}
\phih_{;\nu} + \left(2\xi - \frac{1}{2}\right) g_{\mu\nu}
g^{\rho\sigma} \phih_{;\rho} \phih_{;\sigma} -
2 \xi \phih_{;\mu\nu} \phih \nonumber \\ && +
2 \xi g_{\mu\nu} \phih \square \phih - \xi G_{\mu\nu} \phih^2
+ \frac{1}{2} g_{\mu\nu} \left( m^2 + \frac{\lambda}{4}\phih^2\right)
\phih^2 \Biggr]
\label{eq-onemtcl} 
\end{eqnarray} 
and the quantum part of $\langle T_{\mu\nu} \rangle$ by
\begin{eqnarray}
T^{\text{{\tiny Q}}}_{\mu\nu} = N \hbar \lim_{x' \rightarrow x} \Biggl\{
\Biggl[ &&
(1 - 2\xi) \nabla_{\mu} \nabla_{\nu}' + \left( 2 \xi - \frac{1}{2} \right)
g_{\mu\nu} g^{\rho\sigma} \nabla_{\rho} \nabla_{\sigma}' 
-2\xi \nabla_{\mu} \nabla_{\nu} + 2 \xi g_{\mu\nu}g^{\rho\sigma}
\nabla_{\rho} \nabla_{\sigma}  \nonumber 
\\ && - \xi G_{\mu\nu} + 
\frac{1}{2} g_{\mu\nu} \left( m^2 + \frac{\lambda}{2} \phih^2 +
\frac{\hbar\lambda}{4} G_{++}(x,x') \right) \Biggr] G_{++}(x,x') 
\Biggr\} + O(1).
\label{eq-onemtqm}
\end{eqnarray}
The above expression for $T^{\text{{\tiny Q}}}_{\mu\nu}$ is divergent
in four spacetime dimensions, and needs to be regularized or renormalized.
The energy-momentum
tensor in the one-loop approximation is obtained by neglecting the
$O(\hbar^2)$ terms in Eq.~(\ref{eq-onemtqm}).  It can be shown
using (\ref{eq-onige}) that the energy-momentum tensor 
at leading order in the $1/N$ expansion is covariantly conserved, up to
terms of order $O(1)$ (next-to-leading-order).  The bare
semiclassical Einstein equation is then given (in terms of $\langle
T_{\mu\nu}\rangle$ shown above) by Eq.~(\ref{eq-lpfcstsee2}).

At this point we formally set $N = 1$ since we are not including 
next-to-leading-order diagrams in the $1/N$ expansion.  This can
be envisioned as a simple rescaling of the Planck mass by $\sqrt{N}$, since
the matter field effective action $\Gamma$ is entirely $O(N)$.
We now turn to the issue of renormalization.

\subsection{Renormalization}
\label{sec-onrenorm}
To renormalize the leading-order, large-$N$, CTP effective action in a general 
curved spacetime, one can use dimensional regularization \cite{thooft:1972a}, 
which requires formulating effective action in $n$ spacetime 
dimensions.  This necessitates the introduction of a length parameter 
$\mu^{-1}$ into the classical action, $\lambda \rightarrow \lambda \mu^{4-n}$,
in order for the classical action to have consistent units.  
As above, we maintain the restriction 
$g^{\mu\nu}_{+} = g^{\mu\nu}_{-} = g^{\mu\nu}$, and we suppress indices inside
functional arguments.

Making a substitution of the gap equation into the leading-order, large-$N$,
2PI effective action, we obtain
\begin{eqnarray}
\Gamma[\phih,g^{\mu\nu}] = {\mathcal S}^{\text{{\tiny F}}}[\phih,
g^{\mu\nu}] + \frac{i\hbar N}{2} \text{tr}\,&&\text{ln} \left[
\left( G^{-1} \right)^{ab} \right] \nonumber \\ && +
\frac{\hbar^2 N \lambda \mu^{4-n}}{8}
\int_M d^{\,n}x \sqrt{-g} c^{abcd} \left[ G_{ab}(x,x) G_{cd}(x,x) \right],
\label{eq-eaphi}
\end{eqnarray}
in terms of the CTP propagator $G_{ab}(x,x')$ which satisfies the gap equation
\begin{equation}
\left( G^{-1} \right)^{ab} = i \left( \square_x c^{ab} + \chi^{ab}(x) \right)
\delta(x-x')\frac{1}{\sqrt{-g'}},
\end{equation}
in terms of a four-component ``effective mass''
\begin{equation}
\chi^{ab}(x) = \left( m^2 + \xi R \right) c^{ab} +
\frac{\lambda \mu^{4-n}}{2} c^{abcd} [ \phih_c \phih_d + \hbar
G_{cd}(x,x) ].
\end{equation}
The divergences in the effective action can be made explicit with the
use of the heat kernel $K^a_{\;\;b}(x,y;s)$ 
\cite{dewitt:1965a,thooft:1972a,dowker:1976a}. Let us define 
$K^a_{\;\;b}(x,y,s)$ 
which satisfies
\begin{equation}
\frac{\partial K^a_{\;\;b}(x,y;s)}{\partial s} + \int_M d^{\,n}z
\sqrt{-g_z} c_{cd} (G^{-1})^{ac}(x,z) K^d_{\;\;b}(z,y;s) = 0,
\label{eq-hk}
\end{equation}
with boundary conditions
\begin{equation}
K^a_{\;\;b}(x,y;0) = \delta^a_{\;\;b} \delta(x-y)\frac{1}{\sqrt{-g_y}}
\end{equation}
at $s = 0$ \cite{toms:1982a}.  From Eqs.~(\ref{eq-hk}) and 
(\ref{eq-eaphi}) it follows that
$K^{+}_{\;\;-} = K^{-}_{\;\;+} = 0$ for all $x,$ $y,$ and $s,$ and that
$K^{+}_{\;\;+}$ ($K^{-}_{\;\;-}$) is a functional of $\phih_{+}$ ($\phih_{-}$)
only.  The CTP effective action can then be expressed as 
\begin{equation}
\Gamma[\phih,g^{\mu\nu}] = \Gamma^{+}_{\text{{\tiny IO}}}[\phih_{+},g^{\mu\nu}]
- \Gamma^{-}_{\text{{\tiny IO}}}[\phih_{-},g^{\mu\nu}], 
\end{equation}
in terms of a functional $\Gamma_{\text{{\tiny IO}}}$ on $M$ defined by
\begin{eqnarray}
\Gamma^{+}_{\text{{\tiny IO}}}[\phih_{+},g^{\mu\nu}] = S^{\text{{\tiny F}}}[
\phih_{+},g^{\mu\nu}] - && \frac{i \hbar N}{2} \int_M d^{\,n}x \sqrt{-g}
\int_0^{\infty} \frac{ds}{s} K^{+}_{\;\;+}(x,x;s) \nonumber \\ &&
+ \frac{\hbar^2 N \lambda \mu^{4-n}}{8} \int_M d^{\,n}x \sqrt{-g} 
\left[ \int_0^{\infty} ds 
K^{+}_{\;\;+}(x,x;s)\right]^2,
\end{eqnarray}
and similarly for $\Gamma^{-}_{\text{{\tiny IO}}}$.  It follows from
Eq.~(\ref{eq-hk}) that $K^{+}_{\;\;+}(x,x;s)[\phih_{+}]$ is exactly the same
functional of $\phih_{+}$ as $K^{-}_{\;\;-}(x,x;s)[\phih_{-}]$ is of 
$\phih_{-}$; we denote it by $K(x,x;s)[\phih]$, where $\phih$ is a function
on $M$.

The divergences in the effective action 
arise in the small-$s$ part of the integrations, so that in the equation
\begin{equation}
\int_0^{\infty} \frac{ds}{s} K(x,x;s)= \int_0^{s_0} \frac{ds}{s} K(x,x;s) + 
\int_{s_0}^{\infty} \frac{ds}{s} K(x,x;s)
\end{equation}
only the first term on the right-hand side is divergent.  Using the
$s \rightarrow 0^{+}$ asymptotic expansion for $K(x,x;s)$ \cite{toms:1982a},
one has (for a scalar field, such as the unbroken symmetry, large-$N$ limit
of the O$(N)$ model)
\begin{equation}
K(x,x;s) \sim (4\pi s)^{-\frac{n}{2}} \sum_{m=0}^{\infty} s^m a_m(x),
\end{equation}
where the $a_n(x)$ are the well-known ``Hamidew coefficients'' 
made up of scalar invariants of the spacetime curvature
\cite{dewitt:1975a,dewitt:1965a}.
The divergences then show up as poles in $1/(n-4)$ after the $s$ integrations
are performed.  They have been evaluated for the $\lambda \Phi^4$ theory in 
a general spacetime by many authors (see, e.g., 
\cite{toms:1982a,hu:1984a,paz:1988a}) and in the large-$N$ limit of
the O$(N)$ model \cite{mazzitelli:1989b}.
At leading order in the $1/N$ expansion, the renormalization of 
$\lambda$, $\xi$, $m$, $G$, $\Lambda$, $b$, and $c$ is required,
but no field amplitude renormalization is required.  

\section{Acknowledgments}
This work is supported in part by NSF Grant No.\ 
PHY94-21849.  Part of this work
was carried out at the Institute for Advanced Study, Princeton, where B.L.H.
was a Dyson Visiting Professor.
We enjoyed the hospitality of Los Alamos National Laboratory during the
Santa Fe workshop on {\em Nonequilibrium Phase Transitions\/} sponsored by
the Center for Nonlinear Studies, and organized by Dr.~E.~Mottola. We thank
Dr.~E.~Calzetta for discussions.


\newpage

\end{document}